\documentclass[10pt,a4paper,twoside]{article}
\makeatletter
\usepackage{amsmath}
\usepackage{amsopn}
\usepackage{amssymb}
\usepackage{amstext}
\usepackage{calc}
\usepackage{cite}
\usepackage[usenames]{color}
\usepackage{dsfont}
\usepackage[thinlines]{easymat}
\usepackage{epsfig,epic}
\usepackage{fancyhdr}
\usepackage{rotating}
\usepackage{txfonts}
\usepackage{theorem}
\usepackage[latin1]{inputenc}
\usepackage[T1]{fontenc}
\usepackage[english]{babel}
\newlength{\rilegatura}
\newlength{\simmetricspaceh}
\newcommand{\newatop}[2]{\genfrac{}{}{0pt}{}{#1}{#2}}
\newcommand{\pt}[1]{\left( #1 \right)}
\newcommand{\pq}[1]{\left[ #1 \right]}
\newcommand{\pg}[1]{\left\{ #1 \right\}}
\newcommand{\bs}[1]{\boldsymbol{#1}}
\newcommand{\abs}[1]{\left|\left. #1 \right.\right|}

\newcommand{\norm}[2]{{\left\Arrowvert #1 \right\Arrowvert}_{#2}}
\newcommand{\ud}{\mathrm{d}}

\renewcommand{\Pr}{\mathrm{Pr}}
\newcommand{\tr}{\mathrm{Tr}}
\newcommand{\co}[1]{\textsf{#1}}
\newcommand{\nullla}[1]{#1}
\newcommand{\nulllb}[1]{\overline{#1}}

\newcommand{\coleq}{\coloneqq}
\newcommand{\eqcol}{\eqqcolon}
\renewcommand{\geq}{\geqslant}
\renewcommand{\leq}{\leqslant}
\newcommand{\longpage}{\enlargethispage{2\baselineskip}}

\def\Id{\mathds{1}}

\def\IR{\mathds{R}}

{\theoremstyle{break} \theorembodyfont{\rmfamily}

}
{\theoremstyle{break} \theorembodyfont{\rmfamily}

}
{\theoremstyle{break} \theorembodyfont{\rmfamily}
}
{\theoremstyle{break} \theorembodyfont{\rmfamily}
\newtheorem{AAA}{Algorithm}}
{\theoremstyle{break} \theorembodyfont{\rmfamily}
}
{\theoremstyle{break} \theorembodyfont{\rmfamily}

}
{\theoremstyle{plain}
}

\renewcommand\section{\@startsection{section}{1}{\z@}
{-3.5ex \@plus -1ex \@minus -.2ex}{2.3ex \@plus.2ex}{\normalfont\Large\bfseries\boldmath}}
\renewcommand\subsection{\@startsection{subsection}{2}{\z@}
{-3.25ex\@plus -1ex \@minus -.2ex}{1.5ex \@plus .2ex}{\normalfont\large\bfseries\boldmath}}
\renewcommand\subsubsection{\@startsection{subsubsection}{3}{\z@}
{-3.25ex\@plus -1ex \@minus -.2ex}{1.5ex \@plus .2ex}{\normalfont\normalsize\bfseries\boldmath}}
\renewcommand\paragraph{\@startsection{paragraph}{4}{\z@}
{3.25ex \@plus1ex \@minus.2ex}{-1em}{\normalfont\normalsize\bfseries\boldmath}}
\renewcommand\subparagraph{\@startsection{subparagraph}{5}{\parindent}
{3.25ex \@plus1ex \@minus.2ex}{-1em}{\normalfont\normalsize\bfseries\boldmath}}
\newenvironment{Ventry}[1]
    {\begin{list}{}{
        \settowidth{\labelwidth}{#1}
        \setlength{\leftmargin}{\labelwidth}}}
    {\end{list}}

\definecolor{cyan}{cmyk}{1,0.1,0,0.2}
\definecolor{grey}{rgb}{0.7,0.7,0.7}

\makeatother
\begin{document}

\pagestyle{fancy}
  \lhead[\fancyplain{}{\small \thepage \protect\hspace{5mm}
  {\it Valerio Cappellini, Hans-J{\"u}rgen Sommers, Wojciech Bruzda and Karol {\.Z}yczkowski}}]
       {\fancyplain{}{}}
 \rhead[\fancyplain{}{}]
       {\fancyplain{}{\small {\it Random bistochastic matrices} \hspace{5mm}\thepage}}
 \chead{}\lfoot{}\cfoot{}\rfoot{}
\title{\bfseries \LARGE Random bistochastic matrices}
\author
 {\large Valerio Cappellini$^{1}$, Hans-J{\"u}rgen Sommers$^{2}$, Wojciech Bruzda$^{3}$ and Karol {\.Z}yczkowski$^{3,4}$\\[2ex]
 {\normalsize\itshape $^1$``Mark Kac'' Complex Systems Research Centre, Uniwersytet Jagiello{\'n}ski,
                   ul. Reymonta 4, 30-059 Krak{\'o}w, Poland}\\
 {\normalsize\itshape $^2$Fachbereich Physik, Universit\"{a}t Duisburg-Essen, Campus Duisburg, 47048 Duisburg, Germany}\\
 {\normalsize\itshape $^3$Instytut Fizyki im. Smoluchowskiego, Uniwersytet Jagiello{\'n}ski, ul. Reymonta 4, 30-059 Krak{\'o}w, Poland}\\
 {\normalsize\itshape $^4$Centrum Fizyki Teoretycznej, Polska Akademia Nauk,  Al. Lotnik{\'o}w 32/44, 02-668 Warszawa, Poland}\\[2ex]}
\date{\normalsize (Dated: March 02, 2009)}

\maketitle

\begin{abstract}
\normalsize\phantom{bla}
Ensembles of random stochastic and bistochastic matrices are investigated.
While all columns of a random stochastic matrix can be chosen independently,
the rows and columns of a bistochastic matrix have to be correlated.
We evaluate the probability measure induced into the Birkhoff polytope
of bistochastic matrices by applying the Sinkhorn algorithm to
a given ensemble of random stochastic matrices.
For matrices of order $N=2$ we derive explicit formulae
for the probability distributions induced by random stochastic
matrices with columns distributed according to the Dirichlet distribution.
For arbitrary $N$ we construct an initial ensemble of stochastic matrices which allows one to generate
random bistochastic matrices according to a distribution locally flat at the center of the Birkhoff polytope.
The value of the probability density at this point enables us to obtain an estimation of the volume of the
Birkhoff polytope, consistent with recent asymptotic results.
\\[2.5ex]
\begin{center}
 \noindent{\normalsize e-mails: valerio@ictp.it \ \ \ \ \
                         h.j.sommers@uni-due.de \ \ \ \ \
                             w.bruzda@uj.edu.pl \ \ \ \ \
                               karol@cft.edu.pl}
\end{center}
\end{abstract}

\section{Introduction}\label{intro}

A \emph{stochastic} matrix $M$ is defined as a square matrix of size
$N$, consisting of non--negative elements, such that the sum in each
column is equal to unity. Such matrices provide an important tool often applied
in various fields of theoretical physics, since they represent Markov chains.
In other words, any stochastic matrix maps the set of probability
vectors into itself. Weak positivity of each element of $M$
guarantees that the image vector $p^{\prime}=Mp$ does not contain
any negative components, while the probability is preserved due to the normalization of each
column of $M$.

A stochastic matrix $B$ is called \emph{bistochastic} (or
\emph{doubly stochastic}) if additionally each of its rows sums up
to unity, so that the map preserves identity and for this
reason it is given the name \emph{unital}. Bistochastic matrices
are used in the theory of
\emph{majorization}~\cite{Mar79:1,Bha97:1,CKZ98:1}
and emerge in several physical problems~\cite{B06:1}.
For instance they may represent~a transfer
process at an oriented graph consisting of $N$ nodes.

The set ${\cal B}_N$ of bistochastic matrices of size $N$ can be viewed as~a
convex polyhedron in $\IR^{\pt{N-1}^2}$. Due to the Birkhoff theorem,
any bistochastic matrix can be represented as a convex combination
of permutation matrices. This $(N-1)^2$ dimensional set is often called \emph{Birkhoff polytope}.
Its volume with respect to the Euclidean measure
is known~\cite{Cha99:1,Bec03:1,DeL07:1}  for $2\leq N\leq 10$.

To generate a random stochastic matrix one may take an arbitrary
square matrix with non-negative elements and renormalize each of its
columns. Alternatively, one may generate independently each column
according to a given probability distribution defined on the
probability simplex. A standard choice is the Dirichlet distribution
(\ref{eq14}),  which depends on the real parameter $s>0$
and interpolates between the uniform measure obtained for $s=1$ and the
statistical measure for $s=1/2$ --- see e.g.~\cite{Ben79:1}.

Random bistochastic matrices are more difficult to generate, since
the constraints imposed for the sums in each column and each row imply
inevitable correlations between elements of the entire matrix.
In order to obtain~a bistochastic matrix one needs to normalize all
its rows and columns, and this cannot be performed independently.
However, since the both sets of stochastic and unital matrices are
convex, iterating such a procedure, converges~\cite{BB96} and yields
a bistochastic matrix. Note that initializing the scheme of
alternating projections with different ensembles of initial
conditions leads to various probability measures on the set.

The aim of this work is to analyze probability measures inside the Birkhoff
polytope. In particular we discuss methods of generating random
bistochastic matrices according to the uniform (flat) measure in this set.
Note that the brute force method
of generating random points distributed uniformly inside
 the unit cube of dimension $(N-1)^2$
and checking if the bistochasticity conditions are satisfied,
is not effective even for $N$ of order of $10$,
since the volume of  the Birkhoff polytope ${\cal B}_N$
decreases fast with the matrix size.

The paper is organized as follows. In Section~\ref{gdsmd} we present
after Sinkhorn~\cite{S64:1}
two equivalent algorithms producing~a bistochastic matrix out of any square
matrix of non-negative elements.
An implicit formula (\ref{eq60}) expressing the probability distribution
in the set of bistochastic matrices for arbitrary $N$ is derived in Sec.~3.1,
while exact formulas for the case $N=2$ are presented in Section~\ref{exfor}.
Furthermore, we obtain its power series expansion around the center
$B^{\star}_N$ of the Birkhoff polytope and for each $N$ we single out a
particular initial distribution in the set of stochastic matrices,
such that the output distribution is flat (at least locally) in the
vicinity of $B^{\star}_N$. Finally,  in section 5 we
compute the value of the probability density at this very point and
obtain an estimation of the volume of the set of bistochastic
matrices, consistent with recent results of Canfield and McKay~\cite{Can07:1}.
In Appendix A we demonstrate equivalence of two algorithms used to generate
random bistochastic matrices.
The key expression of this paper (\ref{eq89}) characterising the
probability distribution for random bistochastic matrices
in vicinity of the center of the Birkhoff polytope is
derived in Appendix B, while the third order expansion is worked out in Appendix C.

\section{How to generate a bistochastic matrix?}\label{gdsmd}

\subsection{Algorithm useful for numerical computation}
\begin{figure}[ht]
\begin{center}
\includegraphics[width=2.70in]{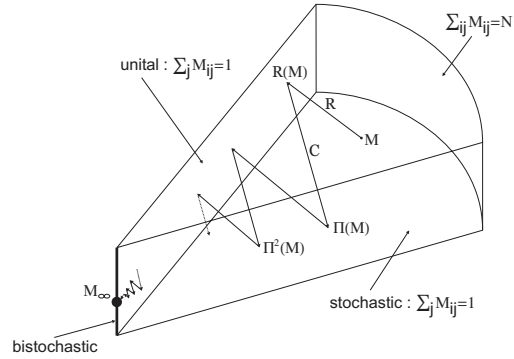}
\caption[]{Sketch of the iteration procedure: a matrix $M$
consisting of non-negative entries is sent by the transformation $R$
(normalization of rows) into the set of unital matrices, and then by
the transformation $C$ (normalization of columns) into the set of
stochastic matrices. Iterating the map $\Pi=(T\circ R)^2$ one arrives at~a
bistochastic matrix $M_{\infty}$.} \label{figW0}
\end{center}
\end{figure}

In 1964 Sinkhorn~\cite{S64:1} introduced the following iterative algorithm
leading to a bistochastic matrix,  based on
alternating normalization of rows and columns of a given square matrix
with non-negative entries:
\begin{quote}
\begin{AAA}[rows/columns normalization]{}\label{alg1}\ \\[-5.5ex]
\begin{Ventry}{\texttt{iii)}}
\item[\texttt{1)}] take an input $N\times N$ stochastic matrix $M$ such that each row
                   contains at least one positive element,
\item[\texttt{2)}] normalize each row-vector of $M$ by dividing it by the sum of its elements,
\item[\texttt{3)}] normalize each column-vector as in the previous point \texttt{2)},
\item[\texttt{4)}] stop if the matrix $M$ is bistochastic up to certain accuracy
                   in some norm $\|\cdot\|$, otherwise go to point \texttt{2)}  .
    \end{Ventry}
    \end{AAA}
    \end{quote}
The above algorithm is symbolically visualized in Fig.~\ref{figW0}.
For an initial point $M$ one may take an arbitrary matrix with
non-negative entries. To fix the scale we may assume that the sum
of all entries is equal to $N$, so $M$ belongs to interior of the $(N^2-1)$
dimensional simplex $\Delta_{N^2-1}$. The transformation $R$ of
normalization of the rows of $M$ produces a unital matrix,
for which the sum of all (non-negative) entries in each row is equal to unity.
 Subsequent normalization of the columns of $R(M)$
maps this matrix into the set of stochastic matrices. This step can
be rewritten as $C=TRT$, where $T$ denotes the transposition of the
matrix. Hence the entire map reads $\Pi\coleq CR=(T\circ R)^2.$ For instance
if $N=2$ in the limit we aim to get a bistochastic matrix
\begin{equation}
\lim_{n\to\infty}\Pi^n\pt{M}\eqcol M_{\infty}=
\begin{pmatrix}
d & 1-d\\
1-d & d
\end{pmatrix}\qquad,\qquad\text{\ for some\ }d\in{\pq{0,1}}.
\label{eq0001}
\end{equation}

Since both these sets are convex, our procedure can be considered
as a particular example of a general construction called
'projections on convex sets'. Due to convexity of these sets the procedure of
alternating projections converges to a point belonging to the intersection of both sets~\cite{BB96}.
An analogous method was recently used by Audenaert and Scheel to generate quantum bistochastic maps~\cite{Aud07:1}.

\subsection{Algorithm suitable for analytical calculation}
To perform analytical calculations of probability distribution
inside the Birkhoff polytope we are going to use yet another
algorithm to generate bistochastic matrix, the idea of which is due
to Djokovi\'{c}~\cite{D70:1}. Already in his earlier
paper~\cite{S64:1} Sinkhorn demonstrated that for a given positive
matrix $M$ there exists exactly one doubly stochastic matrix $B$
such that $B=D^L M D^R$. In order to extend such important result
from posistive matrices to non-negative ones, one has to introduce
the hypotesis of \emph{fully indecomposability}~\cite{S67:1,D70:1}.
For the sake of clarity and reading, we prefer to mention here that
the set of non-fully indecomposable (stochastic) matrices constitute
a zero measure set within the set of all stochastic matrices,
instead of going through the details of Sinkhorn's proof. This means
that the converge of our algorithms we will assume to hold true from
now onwards, has to be intended \emph{almost everywhere} in the
compact set of stochastic matrices, with respect to the usual
Lebesgue measure.

Here $D^L$ and $D^R$ denote diagonal matrices with positive entries
determined uniquely up to~a scalar factor.

To set the notation, we will denote with $\IR^+$ the positive
semi-axis $(0,\infty)$ whereas, the symbol $\IR_+$ will be used for
$\IR^+\cup\pg{0}=[0,\infty)$\,.
Let us now consider the positive cone $\IR_+^{N}$ and the set of
endomorphisms over it, $End\pq{\IR_+^{N}}$, representable by means
of $N\times N$ matrices $M$ consisting of non negative elements $m_{ij}\geq
0$. For any given two vectors $L$ and $R$ in $\IR_+^{N}$, one can
consider a map $\Gamma_{L,R}\in End\pq{End\pq{\IR_+^{N}}}$\,, given
by
\begin{subequations}\label{eq38}
\begin{align}
 End\pq{\IR_+^{N}}\ni M & \longmapsto M^{\prime}=\Gamma_{L,R}\pt{M}\in End\pq{\IR_+^{N}}\label{eq38a}\\
 \IR_+\ni m_{ij}^{\phantom{\prime}} & \longmapsto m^{\prime}_{ij}=\Gamma_{L,R}\pt{m_{ij}^{\phantom{\prime}}}\coleq
 L_{i}^{\phantom{\prime}}\,m_{ij}^{\phantom{\prime}}\,R_{j}^{\phantom{\prime}}\in
 \IR_+\,.\label{eq38b}
\end{align}
\end{subequations}
Defining the positive diagonal matrices
$D^{L}_{ij}\coleq L^{\phantom{L}}_i\,\delta_{ij}^{\phantom{R}}$\;,
and  $D^{R}_{ij}\coleq
R^{\phantom{R}}_i\,\delta_{ij}^{\phantom{R}}$\; respectively, one can observe
that $\Gamma_{L,R}\pt{M}=D^{L}M\,D^{R}$. Our purpose is to design an algorithm that takes a
generic $M\in End\pq{\IR_+^{N}}$ as an input and produces as an output an appropriate pair of
vectors $L,R\in\IR_+^{N}$ such that $\Gamma_{L,R}\pt{M}\eqcol B$ is
bistochastic.

The \emph{stochasticity} condition implies
\begin{subequations}\label{eq39}
 \begin{align}
 \sum_i B_{ij}=1=\sum_i L_{i}\,m_{ij}\,R_{j}\quad\text{\Large
 $\Longrightarrow$}\quad R_{j}>0\quad\text{and}\quad\frac{1}{R_j}=\sum_k L_{k}\,m_{kj}\,.
 \label{eq39a}
 \intertext{Analogously, \emph{unitality} implies}
 \sum_j B_{ij}=1=\sum_j L_{i}\,m_{ij}\,R_{j}\quad\text{\Large
 $\Longrightarrow$}\quad L_{i}>0\quad\text{and}\quad\frac{1}{L_i}=\sum_j m_{ij}\,R_{j}\,,\label{eq39b}
\end{align}
\end{subequations}
so that $L,R\in\pt{\IR^{+}}^{N}\subset\IR_+^{N}$\,.
Both equations~\eqref{eq39} can be merged together into a single equation for
$L$\,,
\begin{equation}
 \frac{1}{L_i}=\sum_j m_{ij}\,\frac{1}{\sum_k L_{k}\,m_{kj}}
\label{eq40}\quad
\end{equation}
which can be interpreted as a kind of \emph{equation of the
motion} for $L$, as it corresponds to a stationary solution of the
\emph{action--like} functional
\begin{equation}
 \Phi\pq{L}=-\sum_i \ln\pt{L_i} + \sum_j \ln\pt{\sum_k
 L_{k}\,m_{kj}}
\label{eq41}\,.
\end{equation}
Equations~(\ref{eq40}--\ref{eq41}) imply that if $L$
is a solution, then for any $\lambda\in\IR$ the rescaled vector $\lambda L$ is as well a solution of~(\ref{eq41}).
Thus we may fix $L_N=1$ and try to
solve~\eqref{eq40} for $L_1,L_2,\ldots,L_{N-1}$\,. Differentiating
eq.~\eqref{eq41} we get
\begin{equation}
 \frac{\partial\Phi}{\partial L_i}=-\frac{1}{L_i}\pq{1-\sum_j
 S_{ij}}\quad,\quad\text{where}\quad S_{ij}\coleq L_i\,m_{ij}\,\frac{1}{\displaystyle\sum_k L_{k}\,m_{kj}}
\label{eq42}
\end{equation}
is a stochastic matrix. Since $L_i\neq 0$, unitality of $S$ is
attained once we impose stationarity to~\eqref{eq42}. Hence the
stationary $L$ implies that $S$ becomes bistochastic. Equation~\eqref{eq41} displays convexity of $\Phi$
 for very small $L_i\ \pt{i=1,2,\ldots,N-1\ ,\ L_N=1}$. The function
 $\Phi$ is convex at
 the stationary point and starts to become concave for large $L_i\,$.
 Thus there is a unique minimum of the
 function $\Phi$\, which can be reached by the following iteration procedure:
\begin{equation}
L_i^{\pt{n}}=\frac{\displaystyle 1}{\displaystyle \sum_j
m_{ij}\,\frac{\displaystyle1}{\displaystyle \sum_k
L_{k}^{\pt{n-1}}\,m_{kj}}} \label{eq45}\,,
\end{equation}
where we fix $L_N$ and iterate the remaining components
$L_1,L_2,\ldots,L_{N-1}$\,only. We start with setting $L_{k}^{\pt{1}}=1\,,\
\forall k$\, which leads to
\begin{quote}
\begin{AAA}[convergent sequences of $\boldsymbol{\IR^N}$ vectors]{}\label{alg4}\ \\[-5.5ex]
\begin{Ventry}{\texttt{iii)}}
\item[\texttt{1)}] take an input $N\times N$ stochastic matrix $M={\pg{m_{ij}}}_{ij}$
                   and define the vector $L^{\pt{0}}={\pt{1,1,\ldots,1}}^{\text{\co{T}}}\in\IR^N$,
\item[\texttt{2)}] run equation~\eqref{eq45} yielding the vector $L^{\pt{n}}$ out of $L^{\pt{n-1}}$,
\item[\texttt{3)}] stop if the matrix $\displaystyle S^{\pt{n}}\coleq L_{i}^{\pt{n}}\,m_{ij}\,
                   \frac{\displaystyle 1}{\displaystyle\sum_k L_{k}^{\pt{n}}\,m_{kj}}$ is
                   bistochastic up to a certain accuracy in some norm $\|\cdot\|$, otherwise
           go to point \texttt{2)}.
\end{Ventry}
\end{AAA}
\end{quote}

The Algorithm (1) is expected to converge faster than the Algorithm (2),
so it can be recommended for numerical implementation.
On the other hand Algorithm (2)
is useful to evaluate analytically the probability measure induced into the Birkhoff polytope
by a given choice of the input ensemble,
and it is used for this purpose in further sections.
The equivalence of these two algorithms is shown in Appendix A.

\section{Probability measures in the Birkhoff polytope}\label{distB}

Assume that the algorithm is initiated with a random matrix $M$ drawn according to a given
distribution $W\big[\{m_{ij}\}\big]$ of
matrices of non negative elements $m_{ij}\geq 0$. We want to
know the distribution of the resulting bistochastic matrices $B_{ij}$ obtained as output
of the Algorithm~(2). To this end, using eq.~\eqref{eq40} and imposing stationarity condition~\eqref{eq42},
we write the distribution for $B$ by integrating over delta functions
\begin{multline}
 P\big[\{B_{ij}\}\big]=
 \int_0^\infty\!\!\!\!\!\cdots\!\!\int_0^\infty\pt{\prod_{r=1}^N\ud L_r}
 \int_0^\infty\!\!\!\!\!\cdots\!\!\int_0^\infty\pt{\prod_{p,q=1}^N\ud m_{pq}\;W\big[\{m_{pq}\}\big]}
 \;\times\prod_{i,j=1}^N\delta\pt{B_{ij}- L_i\,m_{ij}\,\frac{1}{\displaystyle\sum_k
 L_{k}\,m_{kj}}}\;\times\\
 \times\delta\pt{L_N-1}\prod_{u=1}^{N-1}\delta\pt{-\frac{1}{L_u}+\sum_t
 m_{ut}\,\frac{1}{\displaystyle\sum_v L_{v}\,m_{vt}}}
 \;\times J\pg{L_1,L_2,\ldots,L_{N-1}}\,,\label{eq53}
\end{multline}
where the Jacobian factor reads
\begin{equation}
 J\pg{L_1,L_2,\ldots,L_{N-1}}\coleq
 \det\pq{\frac{\partial^2\Phi}{\partial L_i\,\partial
 L_\ell}}_{i,\ell=1}^{N-1}
=
 \pt{\prod_{i=1}^{N-1}\frac{1}{L_i^2}}\;\times\det{\pq{\Id-BB^{\text{\co{T}}}}}_{N-1}
 \label{eq54}\,.
\end{equation}

\noindent Here and in the following
${\pq{\Id-BB^{\text{\co{T}}}}}_{N-1}$\, will indicate the
$\pt{N-1}\times\pt{N-1}$ block matrix $\pq{\delta_{i\ell}
 -\sum_{j=1}^N B_{ij}\:B_{\ell j}}_{i,\ell=1}^{N-1}$\,, that is positive defined, and
the symbol $P\big[\{A_{ij}\}\big]$\, will denote the
probability density $P$ of matrices $A=\{A_{ij}\}$. This notation
will also be used for matrices whose elements are functions of
elements of another matrix, namely $P\big[\{f(A_{ij})\}\big]$\,.
Plugging eq.~\eqref{eq54} into~\eqref{eq53} and introducing again
the delta functions for variables $R_j$ of~\eqref{eq39a} we obtain
\begin{align}
 P\big[\{B_{ij}\}\big]&=
 \int_0^\infty\!\!\!\!\!\cdots\!\!\int_0^\infty\pt{\prod_{r=1}^N\ud L_r}
 \int_0^\infty\!\!\!\!\!\cdots\!\!\int_0^\infty\pt{\prod_{s=1}^N\ud R_s}\delta\pt{L_N-1}
 \int_0^\infty\!\!\!\!\!\cdots\!\!\int_0^\infty\pt{\prod_{p,q=1}^N\ud m_{pq}\;W\big[\{m_{pq}\}\big]}
 \;\times\prod_{i,j=1}^N\delta\pt{B_{ij}-
 L_i\,m_{ij}\,R_j}\;\times\nonumber\\
 &\qquad\times\prod_{u=1}^{N-1}\delta\pt{-\frac{1}{L_u}+\sum_t
 m_{ut}\,R_t}\;\times
 \prod_{w=1}^{N}\delta\pt{R_w-\frac{1}{\displaystyle\sum_h L_{h}\,m_{hw}}}
 \;\times
 \prod_{z=1}^{N-1}\frac{1}{L_z^2}\;\times\det{\pq{\Id-BB^{\text{\co{T}}}}}_{N-1}\,.\label{eq55}
 \intertext{Using the property of the Dirac delta function and
 making use of the Heaviside step function $\theta$, we perform integration over the variables $\ud m_{pq}$.
Introducing new variables $\alpha_i\coleq 1/L_i$ and $\beta_i\coleq
 1/R_i$\,, so that $\ud L_i\;\ud R_j\mapsto\displaystyle L_i^2\;R_j^2\;\ud\alpha_i\;\ud \beta_j$\,, we get}
 P\big[\{B_{ij}\}\big]&=
 \int_0^\infty\!\!\!\!\!\cdots\!\!\int_0^\infty\pt{\prod_{r=1}^N\ud \alpha_r\;\alpha_r^{\,N-1}}
 \int_0^\infty\!\!\!\!\!\cdots\!\!\int_0^\infty\pt{\prod_{s=1}^N\ud \beta_s\;\beta_s^{\,N-1}}
 \prod_{p,q=1}^N W\pq{\pg{\alpha_p B_{pq}\,\beta_q}}\;\delta\pt{\alpha_N-1}\;\times\nonumber\\
 &\qquad\times\det{\pq{\Id-BB^{\text{\co{T}}}}}_{N-1}\;\times\prod_{u=1}^{N-1}\delta\pt{1-\sum_t
 B_{ut}}\;\times
 \prod_{w=1}^{N}\delta\pt{1-\displaystyle\sum_h B_{hw}}\;\times\prod_{a,c=1}^N\theta\pt{B_{ac}}
 \,.\label{eq56}
\end{align}
 The last three factors show that $B_{ij}$ is bistochastic. The factor
 $\det{\pq{\Id-BB^{\text{\co{T}}}}}_{N-1}$ indicates that the expression is meaningful
 only in the case for which the leading eigenvalue $1$ of $BB^{\text{\co{T}}}$ is non-degenerate.

 If the matrix $m_{ij}$ is already stochastic,
 \begin{equation}
 W\big[\{m_{pq}\}\big]= V\big[\{m_{pq}\}\big]\;\times\prod_{w=1}^{N}\;\delta\pt{1-\displaystyle\sum_h
 m_{hw}}\;\times\prod_{a,c=1}^N\theta\pt{m_{ac}}\,,\label{eq57}
 \end{equation}
 then the integration over $\beta_j$ can be performed and we arrive at the final expression
 for the probability distribution inside the Birkhoff polytope which depends on the initial measure
 $V$ in the set of stochastic matrices;
\begin{align}
 P\big[\{B_{ij}\}\big]&=
 \int_0^\infty\!\!\!\!\!\cdots\!\!\int_0^\infty\pt{\prod_{r=1}^N\ud
 \alpha_r\;\alpha_r^{\,N-1}}\;
 \prod_{t=1}^N \;\frac{1}{\pt{\sum_{s}\alpha_s B_{st}}^{\,N}}\;
 \prod_{p,q=1}^N
 V\pq{\pg{\alpha_p B_{pq}\;\frac{1}{\sum_{r}\alpha_r B_{rq}}
 }}\;\delta\pt{\alpha_N-1}\;\times\nonumber\\
 &\qquad\times\det{\pq{\Id-BB^{\text{\co{T}}}}}_{N-1}\;\times\prod_{u=1}^{N-1}\delta\pt{1-\sum_t
 B_{ut}}\;\times
 \prod_{w=1}^{N}\delta\pt{1-\displaystyle\sum_h B_{hw}}\;\times\prod_{a,c=1}^N\theta\pt{B_{ac}}
 \,.
\label{eq60}
\end{align}
The above implicit formula, valid for any matrix size $N$
and an arbitrary initial distribution $V$,
constitutes one of the key results of this paper. It will be
now used to yield explicit expressions
for the probability distribution inside the set of bistochastic matrices
for various particular cases of the problem.

\subsection{Measure induced by Dirichlet distribution}

Let us now assume that the initial stochastic matrices are formed of $N$ independent
columns each distributed according to the Dirichlet distribution~\cite{Sla99:1,Zyk01:1,Ben79:1},
\begin{equation}
D_s(\lambda_1^{\phantom{1}},\ldots,\lambda_{N-1}^{\phantom{1}}) =
\alpha_s \;\lambda_1^{s-1}\ldots\;
\lambda_{N-1}^{s-1}(1-\lambda_1^{\phantom{1}}-\cdots-\lambda_{N-1}^{\phantom{1}})^{s-1}\,,
\label{eq14}
\end{equation}
where $s > 0$ is a free parameter and the normalization constant reads
$\alpha_s=\Gamma\pq{2s}/\Gamma\pq{s}^2$.

\begin{quote}
\begin{AAA}[Random points in the simplex according to the Dirichlet distribution]{}\label{alg2}\ \\[-2ex]
    Following~\cite{Dev86:1} we are going to sketch here a
      useful algorithm for generating random points in a simplex $\boldsymbol{\Delta}_{N-1}$ according to the
    distribution~\eqref{eq14}.
\begin{Ventry}{\texttt{ii)}}
\item[\texttt{1)}] generate an $N$--dimensional vector $X$, whose elements are independent
        random numbers $x_{i}$ from the gamma
        distribution $f\pt{x_i;s,1}$ of \emph{shape} $s$ and \emph{rate} $1$, so that each of them is drawn according to the probability
        density $x_i^{s-1}\mathrm{e}^{-x_i}/\Gamma\pt{s}$\ ;
\item[\texttt{2)}] normalize the vector $X$ by dividing it by its
        $\ell_1$ norm, $X\longmapsto Y\coleq X/\norm{X}{1}$,
        so that the entries will become $x_i\longmapsto y_i\coleq x_i/\sum_{k=1}^N
        x_k$\ .
\end{Ventry}
\end{AAA}
\end{quote}
A simplified version, suited for (semi)integer $s$ is described in
the appendix of~\cite{Zyk01:1}. In particular, to get the uniform
distribution in the simplex $(s=1)$\,, it is sufficient to generate
$N$ independent complex Gaussian variables (with mean zero and
variance equal to unity) and set the probability vector by
\begin{equation}
y_i=|z_i|^2/\sum_{i=1}^N|z_i|^2. \label{eq16F}
\end{equation}
\noindent

Hence the initial stochastic matrix $M$ is characterized by
the vector consisting of $N$ Dirichlet parameters ${\bs{s}}=\{s_1,
\dots , s_N\}$, which determine the distribution of each column.

The probability density can be written as
\begin{equation}
V_{\bs{s}}\big[\{m_{ij}\}\big]\coleq\prod_{j}
D_{s_j}\pt{m_{1j}\,,\,m_{2j}\,,\,\ldots\,,\,m_{N-1j}}= {\cal N}\;
\prod_{ij}{\pt{m_{ij}}}^{s_j-1}  \ ,
\label{eq61-62}
\end{equation}
where the normalisation factor reads
\begin{equation}
{\cal N}= \prod_{j=1}^N\;
\frac{\Gamma\pt{Ns_j}}{{}^{\phantom{N}}\Gamma\pt{s_j}^N}
\label{eq62}\,.
\end{equation}
\noindent
Thus one can obtain the probability distribution of the product
\begin{align}
 V_{\bs{s}}\big[\{\alpha_p B_{pq}\,\beta_q\}\big]& = {\cal N}\; \prod_{pq}{\pt{\alpha_p B_{pq}\,\beta_q}}^{s_q-1}
 = {\cal N}\; \prod_{pq}{B_{pq}}^{s_q-1}\;\times
 \prod_{x}{\alpha_x }^{\sum_y s_y-N}\;\times
 \prod_{z}{\beta_z}^{N\pt{s_z-1}}
 \intertext{and making use of eq. (\ref{eq60})
one eventually arrives at a compact expression for the probability distribution
   in the set of bistochastic matrices}
 P_{\bs{s}}\big[\{B_{ij}\}\big]&=
 {\cal N}\;\int_0^\infty\!\!\!\!\!\cdots\!\!\int_0^\infty\pt{\prod_{r=1}^N\ud
 \alpha_r\;\alpha_r^{\sum_y s_y-1}}\;
 \prod_{t=1}^N \;\frac{1}{\pt{\sum_{j}\alpha_j B_{jt}}^{\,Ns_t}}\;\,
 \delta\pt{\alpha_N-1}\;\times\,\prod_{p,q=1}^N
 {B_{pq}}^{s_q-1}\;\times\nonumber\\
 &\qquad\times\,\times\,\prod_{u=1}^{N-1}\delta\pt{1-\sum_t
 B_{ut}}\;\times
 \prod_{w=1}^{N}\delta\pt{1-\displaystyle\sum_h B_{hw}}\;\times\prod_{a,c=1}^N\theta\pt{B_{ac}}
 \,.\label{eq63}
\end{align}
\noindent Although the results were obtained under the assumption
that the initially random stochastic matrices are characterized by
the Dirichlet distributions~(\ref{eq61-62},\ref{eq62}), one may also
derive analogous results for other initial distributions. As
interesting examples, one can consider the one--parameter family
$V_{\bs{s}\,,\,\lambda\,}\big[\{m_{ij}\}\big]$, in which each
    $j$--column of $M$ is drawn according to a different gamma
    distribution $f\pt{m_{ij};s_j,\lambda}$ of \emph{shape} $s_j$ and \emph{rate}
    $\lambda$\,, that is
\begin{equation}
V_{\bs{s}\,,\,\lambda\,}\big[\{m_{ij}\}\big]= \prod_{j=1}^N\;
\frac{\lambda^{Ns_j}}{{}^{\phantom{N}}\Gamma\pt{s_j}^N}\;
\prod_{ij}\mathrm{e}^{-\lambda
m_{ij}}\;{\pt{m_{ij}}}^{s_j-1}\label{eq67a}
\end{equation}
or, allowing the exponents $s$ to vary through the whole matrix, we can start with
\begin{equation}
\quad V_{\{s_{ij}\}\,,\,\lambda\,}\big[\{m_{ij}\}\big]=
\prod_{ij}\pq{\mathrm{e}^{-\lambda
m_{ij}}\;{\pt{m_{ij}}}^{s_{ij}-1}}
\frac{\lambda^{s_{ij}}}{{}\Gamma\pt{s_{ij}}}\,.\label{eq67b}
\end{equation}
and recover~\eqref{eq63}\,, independently on the rate
$\lambda$ labeling the input.

\subsection{Probability measures for $N=2$}\label{exfor}

\begin{figure}[ht]
\begin{center}
\includegraphics[width=\textwidth]{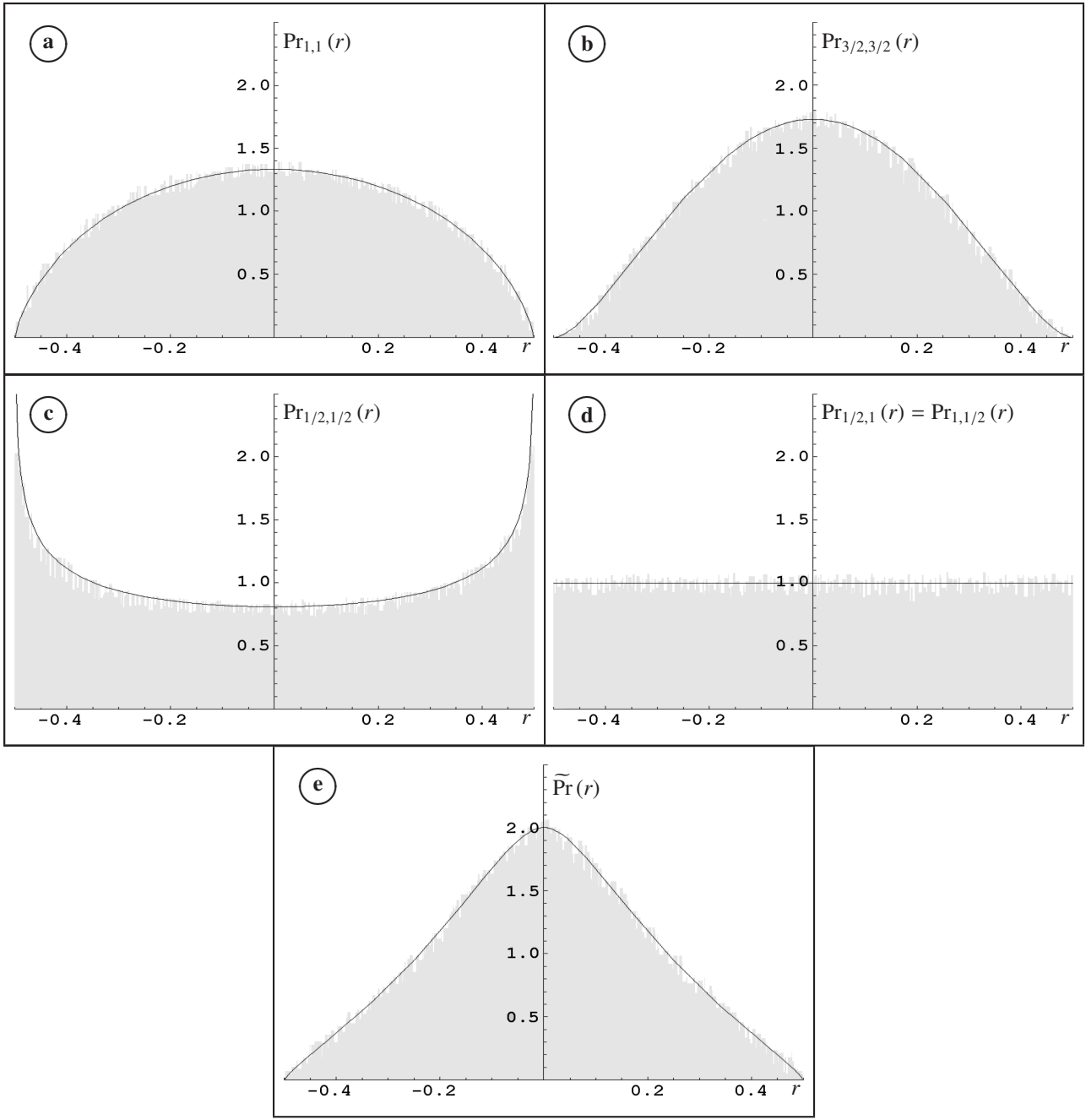}
\caption{
Probability distribution $\Pr(r)$ in the set of $N=2$ bistochastic matrices for
various initial measures. Histograms obtained numerically for a sample of $10^6$ initial matrices
by applying Algorithm~(1) are compared with analytical probability distributions (solid lines);
{\bfseries (a)} semicircle--like~(\ref{eq223}) for $\Pr_{1, 1}$;
{\bfseries (b)} Gaussian--like~(\ref{eq224}) for $\Pr_{3/2, 3/2}$;
{\bfseries (c)} convex distribution~(\ref{eq225}) for $\Pr_{1/2, 1/2}$;
{\bfseries (d)} flat distribution~(\ref{eq226}) for $\Pr_{1/2, 1}$ and
{\bfseries (e)} distribution~(\ref{eq226_6}).}\label{figtwo}
\end{center}
\end{figure}

In the simplest case, for $N=2$ and $B_{ij}=\begin{pmatrix}
                                       d & 1-d \\
                                       1-d & d
                                     \end{pmatrix}$\,,
formula~(\ref{eq63}) describes the probability measure
$P_{s_1,s_2}(d)$ induced into the set of bistochastic matrices by
the ensemble of stochastic matrices with two independent columns
distributed according to the Dirichlet measure with parameters $s_1$
and $s_2$; after integration on $\alpha_2$, renaming
$\alpha_1$ into $\alpha$, and expressing
$\det{\pq{\Id-BB^{\text{\co{T}}}}}_{N-1}=\;2\,d\pt{1-d}$, we arrive
at
\begin{equation}
 P_{s_1,s_2}(d)=
 {\cal N}\Big|_{N=2}\;\int_0^\infty\ud
 \alpha\;\alpha^{s_1+s_2-1}
 \pq{\frac{1}{\alpha d+1-d}}^{\,2s_1}\pq{\frac{1}
{\alpha \pt{1-d}+d}}^{\,2s_2}2\,{\pq{d\pt{1-d}}}^{s_1+s_2-1}\,\theta\pt{d}\,\theta\pt{1-d} \ .
 \label{eq64}
\end{equation}
This expression can be explicitly evaluated for exemplary
 pairs of the Dirichlet parameters $s_1$ and $s_2$,
\begin{align}
 \Pr_{1,1}\pt{r}&=\frac{\pt{1-4r^2}\pq{\pt{1+4r^2}\ln\pt{\frac{1+2r}{1-2r}}-4r}}{16r^3}\,,\label{eq223}\\
 \Pr_{3/2,3/2}\pt{r}&=\frac{{\pt{1-4r^2}}^2\pq{\pt{3+8r^2+48r^4}\ln\pt{\frac{1+2r}{1-2r}}-12r-48r^3}}{16\pi^2r^5}\,,\label{eq224}\\
 \Pr_{1/2,1/2}\pt{r}&=\frac{2\ln\pt{\frac{1+2r}{1-2r}}}{\pi^2r}\,,\label{eq225}\\
 \Pr_{1/2,1}\pt{r}&=\Pr_{1,1/2}\pt{r}=1 \ ,
\label{eq226}
\end{align}
where $r=d-\frac{1}{2}$.
These distributions are plotted in Fig.~\ref{figtwo} and compared with the numerical results.

There is another important distribution that we would like to
consider. We started our analysis by considering a stochastic matrix
as an input state of the renormalization algorithm. However, as an
initial point one can also take a generic matrix $K$ whose four
entries $\pg{k_{11},\ k_{12},\ k_{21},\ k_{22}}$ are just uniformly
distributed on some interval. After the first application of the
half--step map $T\circ R$, (see Fig.~\ref{figW0}) as
\begin{equation}K=
\begin{pmatrix}
k_{11}& k_{12}\\
k_{21}& k_{22}
\end{pmatrix}\xrightarrow[]{\quad \displaystyle T\circ R\quad}
\begin{pmatrix}
\displaystyle \frac{k_{11}}{k_{11}+k_{12}} & \displaystyle\frac{k_{21}}{k_{21}+k_{22}}\\
\rule[0pt]{0pt}{5ex}\displaystyle\frac{k_{12}}{k_{11}+k_{12}} &
\displaystyle\frac{k_{22}}{k_{21}+k_{22}}
\end{pmatrix}=
\begin{pmatrix}
a & 1-b\\
1-a& b
\end{pmatrix}\,,
\label{eq26_1}
\end{equation}
matrix $K$ becomes stochastic, so that this problem can be reduced to the
framework developed so far.

The joint probability distribution of $N$ independent random numbers
$y^{\prime}_i$, drawn according to the uniform distribution in
one interval of $\mathds{R}^+$, and then rescaled as
\begin{equation}
y^{\prime}_i \rightarrow
y_i^{\phantom{\prime}}=\frac{y^{\prime}_i}{\sum_{i=1}^N
y^{\prime}_i}\,,\label{eq26_2}
\end{equation}
reads $P(y_1....y_N) =
\delta(1-\sum_iy_i)/\pg{N\pq{\max\pt{y_i}}}^N$ ~\cite{Zyk01:1}. In the simplest
case, $N=2$, it gives $\widetilde{p}(y)=1/2y^2$ for $y\in(1/2,1]$,
(where $y\coleq y_1=1-y_2$) and symmetrically for $y\in\pq{0,1/2}$. Using this and assuming
independence between the entries of the matrix $K$, the distribution for
the variable $a$ and $b$ of~\eqref{eq26_1} reads
\begin{equation}
 \widetilde{P}\pt{a,b}\coleq \widetilde{p}(a) \times
 \widetilde{p}(b)={\pt{\frac{1}{2\:\max\pg{a,1-a}\:\max\pg{b,1-b}}}}^2\,.
 \label{eq26_3}
\end{equation}

\longpage

Plugging the last expression into the r.h.s. of~\eqref{eq64} we
obtain (see Fig.~\ref{figtwo}(e))
\begin{equation}
  \widetilde{\Pr}\pt{d}=
  \rule[0pt]{0pt}{7ex}=\frac{2\pt{1-2\abs{r}}\pq{1+2\ln\pt{\frac{1+2\abs{r}}{1-2\abs{r}}}}}{{\pt{1+2\abs{r}}}^3}
  \label{eq226_6}\,,
\end{equation}
where again $r=d-1/2$.

 \clearpage

\subsection{Symmetries and relations with the unistochastic matrices for $N=2$}\label{srum}

Consider the map $T\circ R$ defined in~\eqref{eq26_1} acting on an initially stochastic matrix
$\left(\begin{array}{cc}a&1-b\\1-a&b\end{array}\right).$ The symmetry of this map
with respect to diagonal lines $a=b$ and $a=-b$ implies that:
\begin{itemize}
    \item the limit distribution $\Pr_{s_a,s_b}\pt{r}$ is an even
 function of $r$,
    \item $\Pr_{s_a,s_b}=\Pr_{s_b,s_a}$, for any $s_a$ and
 $s_b$. The final accumulation point $d\in [0, 1]$
 can be achieved from the point $\pt{a,b}$ as well as from $\pt{b,a}$.
\end{itemize}
In particular the second point implies that \textbf{if}
$\Pr_{s_a,s_b}\pt{r}$ is the output probability density when the
$\pt{a,b}$--distribution is given by $P_{s_a,s_b}\pt{a,b}$ and
\textbf{if} $s_a\neq s_b$ \textbf{then} for any given
$\lambda\in\pq{0,1}$ the distribution
$\lambda\,P_{s_a,s_b}\pt{a,b}+\pt{1-\lambda}\,P_{s_b,s_a}\pt{a,b}$
will give the same output. Using this we can restore the symmetry
between $\pt{a,b}$ simply by picking $\lambda=1/2$.
\begin{equation}P_{\pq{1/2,1}}^{\text{sym}}\pt{a,b}\coleq\frac{1}{2}P_{1/2,1}\pt{a,b}+\frac{1}{2}P_{1,1/2}\pt{a,b}=
 \frac{1}{2\pi\sqrt{a\pt{1-a}}}+ \frac{1}{2\pi\sqrt{b\pt{1-b}}}
 \label{eq27}\quad
\end{equation}
is a symmetric distribution for $a$ and $b$ which produce, at a long
run, the uniform distribution $P(d)=1$.
Note that the above formula is not of a product form,
so the distribution in both columns are correlated.
In fact such a probability distribution can be interpreted
as a classical analogue of the quantum entangled state \cite{Tu00,Ben79:1}.

Random  pairs $\pt{a,b}$ distributed according to distribution~\eqref{eq27} can be
generated by means of the following algorithm,
\begin{quote}
\begin{Ventry}{\texttt{ii)}}
\item[\texttt{1)}] generate the number $a$ according to $D_{1/2}\pt{a}$ and $b$ according to $D_{1}\pt{b}$
\item[\texttt{2)}] flip a coin: on tails do nothing, on heads exchange $a$ with $b$
\end{Ventry}
\end{quote}
For $N=2$ there exists an equivalence between the set of
bistochastic and unistochastic matrices~\cite{Zyk03:1}. The latter
set is defined as the set of $2\times 2$ matrices whose entries
are squared moduli of entries of
 unitary matrices. The Haar measure on $U(2)$ induces a natural,
 uniform measure in the set of unistochastic matrices: if $U$ is random then
 $P(|U_{11}|^2)=P(|U_{22}|^2)=1$ on $[0, 1].$ Hence initiating Algorithm (1)
 with stochastic matrices distributed according to eq.~(\ref{eq27})
 we produce the same measure in the set of bistochastic matrices as it is induced by the Haar measure on $U(2)$
 by the transformation $B_{ij}=|U_{ij}|^2$.

\section{In search for the uniform distribution for an arbitrary $N$}

For an arbitrary $N$ we shall compute the probability density at the center
$B_N^{\star}$ of the Birkhoff polytope,
\begin{equation}
B^{\star}_N=\begin{pmatrix} \rule[-7pt]{0pt}{7pt}
\ \frac{1}{N}&\frac{1}{N}&\dots&\frac{1}{N}\ \\
\ \frac{1}{N}&\frac{1}{N}&\dots&\frac{1}{N}\ \\
\hdotsfor[2]{4}\\
\ \frac{1}{N}&\frac{1}{N}&\dots&\frac{1}{N}\
\end{pmatrix}\,.
\label{eq65bis}
\end{equation}

Let us begin our analysis by expanding $P_{\bs{s}}\big[\{B_{ij}\}\big]$ around
the center $B^{\star}_N$~\eqref{eq65bis} of the Birkhoff polytope. We start from~\eqref{eq63}
with ${\cal N}$ given by equation~\eqref{eq62}\,, so that
\begin{align}
 P_{\bs{s}}\big[\{B_{ij}\}\big]&=\widetilde{P}_{\bs{s}}\big[\{B_{ij}\}\big]\times\,\prod_{u=1}^{N-1}\delta\pt{1-\sum_t
 B_{ut}}\;\times
 \prod_{w=1}^{N}\delta\pt{1-\displaystyle\sum_h B_{hw}}\;\times\prod_{a,c=1}^N\theta\pt{B_{ac}}
 \quad \label{eq70}
 \intertext{and}
 \widetilde{P}_{\bs{s}}\big[\{B_{ij}\}\big]&=
 {\cal N}\;\int_0^\infty\!\!\!\!\!\cdots\!\!\int_0^\infty\pt{\prod_{r=1}^N\ud
 \alpha_r\;\alpha_r^{\sum_y s_y-1}}\;
 \prod_{w=1}^N \;\frac{1}{\pt{\sum_{s}\alpha_s B_{st}}^{\,Ns_w}}\;\,
 \delta\pt{\alpha_N-1}\;\times\nonumber\\
 &\qquad\qquad\qquad\;\times\,\prod_{p,q=1}^N
 {B_{pq}}^{s_q-1}\times\,\det{\pq{\Id-BB^{\text{\co{T}}}}}_{N-1}
 \,,\label{eq71}
\end{align}
on the manifold $\sum_t B_{ut}=\sum_h B_{hw}=1$\,, \ $B_{ac}\geq 0$\,.

\subsection
{Expansion of probability distribution around the center of the polytope}\label{sec4.1}

Expanding $\widetilde{P}_{\bs{s}}\big[\{B_{ij}\}\big]$ in
power of $\delta B_{ij}$ with
\begin{equation}
 B_{ij} = \frac{1}{N} +  \delta B_{ij}\,,\,
 \sum_i \delta B_{ij}=\sum_j \delta B_{ij}=0\,.\label{eq72}
\end{equation}

we obtain, as shown in Appendix~B, the following result
\begin{align}
 \widetilde{P}_{\bs{s}}\big[\{B_{ij}\}\big]
 =P^{\star}_N\left\{1+\;\pt{\frac{N^2}{2}-1}\sum_{pq}{\pt{\delta
B_{pq}}}^2
 -\frac{\sigma\,N^3}{2\pt{\sigma\,N+1}}\:\sum_{p^{\phantom{\prime}}\!\!q} {\,s_q} \,{\pt{\delta
 B_{pq}}}^2
 +\frac{N^3}{2\pt{\sigma N+1}}\sum_{p^{\phantom{\prime}}} \,{\pt{\sum_q
 {\,s_q}\,\delta B_{pq}}}^2
 +
 {\cal O}\pt{\pt{\delta B}^3}\right\}\,,
\label{eq89}
\end{align}
where $\sigma=\sum_{j=1}^N s_j$ denotes the sum of the
Dirichlet parameters for each column and the factor
\begin{equation}
 P^{\star}_N\coleq P_{\bs{s}}\big[\{\ B_{ij}=1/N\ ,\ \forall ij\ \}\big]\ =\ N^{N^2-1}\;
 \frac{\Gamma\pt{\sum_m s_m}^{N}}{\Gamma\pt{N \sum_m s_m}^{\phantom{N}}}
 \prod_{n=1}^N\;
 \frac{\Gamma\pt{Ns_n}}{\Gamma\pt{s_n}^{N}}
 \,,
\label{eq66}
 \end{equation}
is equal to the value of the probability distribution at the center of the
polytope ${\cal B}_N$, which corresponds to $\delta B=0$.

Assume now that there exists a set of Dirichlet exponents $s_i$,
such that $\widetilde{P}_{\bs{s}}\big[\{B_{pq}\}\big]$ is constant
on the required manifold~\eqref{eq72}\,. Then the quadratic form in
$\delta B_{pq}$ must be identically zero. For $N=2$ this yields only
one equation for two exponents, $2 s_1 + 2 s_2 + 1 = 8 s_1 s_2$\,,
which can e.g. be fulfilled by $s_1=1/2$ and $s_2=1$ (compare
with Section~\ref{srum}).

For $N\geq 3$\,, however, this gives more independent equations, in
general $\pt{N-1}^2$, namely the number of independent variables
parameterizing the Birkhoff polytope. Being $N$ the number of
exponents to be determined, if a solution exists, then it is unique.
Actually the solution exists\,, and corresponds to take all $s_i$
equal to each other: let's call $s$ this collective exponent. Within
this constraint, the last term in~\eqref{eq89} drops out, because of
equation~\eqref{eq72}\,, and the entire quadratic form can be zero,
provided that we choose
\begin{equation}
 \pt{\frac{N^2}{2}-1}=\frac{\sigma\,N^3}{2\pt{\sigma\,N+1}}s^{\star}
 \label{eq90}\quad.
\end{equation}
Now, setting $\sigma=N s^{\star}$, we arrive at $N^4
{s^{\star}}^2=\pt{N^2s^{\star}+1}\pt{N^2-1}$, whose unique positive
solution is
\begin{equation}
s^{\star}=\frac{1}{2N^2}\pt{N^2-2+\sqrt{N^4-4}}
 =1-\frac{1}{N^2}-\frac{1}{N^4}+{\cal O}\pt{\frac{1}{N^8}}\,.\label{eq91}
\end{equation}
The distribution generated by the choice $s=s^{\star}$
will be flat at the center of the polytope but it needs not to be globally uniform.

It is not possible to find an initial Dirichlet distribution which gives the output
distribution uniform in the vicinity of the center of the Birkhoff polytope up to the
third order --- see Appendix C.

\subsection{Numerical results for $N=3$}\label{sec4.5}

Properties of the measures induced in the space of bistochastic
matrices by applying the iterative Algorithm (1) were
analyzed for $N=3$\,. As a starting point we took a random stochastic
matrix $M$ generated according to the Dirichlet distribution~(\ref{eq14}) with
the same parameter for all three columns, $s_1=s_2=s_3=s$. The
resulting bistochastic matrix, $B=\lim_{n\to \infty} \Pi^n(M)$, can
be parameterized by
\[
B\ = \ \left[
\begin{array}{ll|l}
  B_{11} & B_{12} & \ast \\
  \rule[-5pt]{0pt}{2pt}B_{21} & B_{22} & \ast \\
  \hline
  \ast & \ast & \ast \\
\end{array}
\right],
\]
where the $\ast$-marked entries depend on the entries $B_{jk}$\,,
with $j,k\in\pg{1,2}$\,. A sample of initial points consisted of
$10^8$ stochastic matrices generated according to the Dirichlet
distribution with the optimal value
$s^{\star}=\frac{1}{18}(7+\sqrt{77})$
which follows from eq.~\eqref{eq91}).
It produces an ensemble covering the entire $4D$ Birkhoff polytope formed
by the convex hull of the six different permutation matrices of
order three.

\begin{figure}[htbp]
\begin{center}
\includegraphics[width=2.70in]{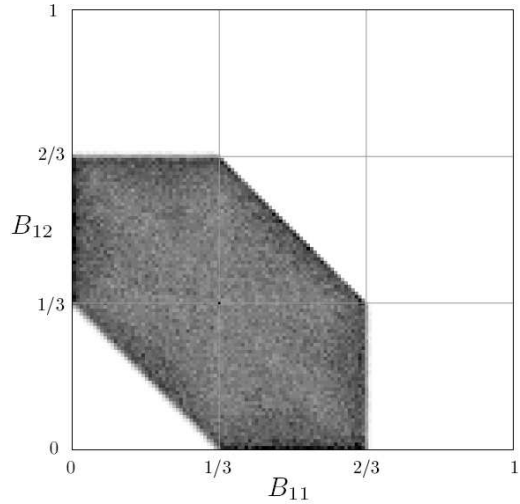}
\caption{Probability density at a subset of the Birkhoff polytope
for $N=3$\,, the ``fat'' hexagon characterized by $\pq{\,B_{11}\,,
\,B_{12}\,,\, \frac{1}{3} \pm 0.01\,,\, \frac{1}{3} \pm 0.01\,}$\,,
for initially stochastic matrices generated with the Dirichlet parameter $s^{\star}$ given by
eq.~\eqref{eq91}\,.}
\label{figW1}
\end{center}
\end{figure}

To visualize numerical results we selected the cases for which
$B_{21}=B_{22}=1/3 \pm 0.01$. Such a  two dimensional cross-section
of the Birkhoff polytope has a shape of a hexagon at the plane
$(B_{11},B_{12})$, centered at the center of the body, $B^{\star}_3=
\pq{\,\frac{1}{3}\,,\, \frac{1}{3}\,, \frac{1}{3}\,; \dots,\, \frac{1}{3}\,}$\,.
Figure~\ref{figW1} shows the probability distribution along this
section, obtained from these $4 \times 10^6$ realizations of the
algorithm which produce bistochastic matrices inside a layer of width
$0.02$ along the section.

As expected for the critical value $s^{\star}$ of the Dirichlet
parameter, the resulting distribution is flat in the vicinity of the
center of the polytope. However, this distribution is not globally
uniform and shows a slight enhancement of the probability  (darker
color) along the boundary of the polytope.

\begin{figure}[ht]
\begin{center}
\includegraphics[width=3.00in]{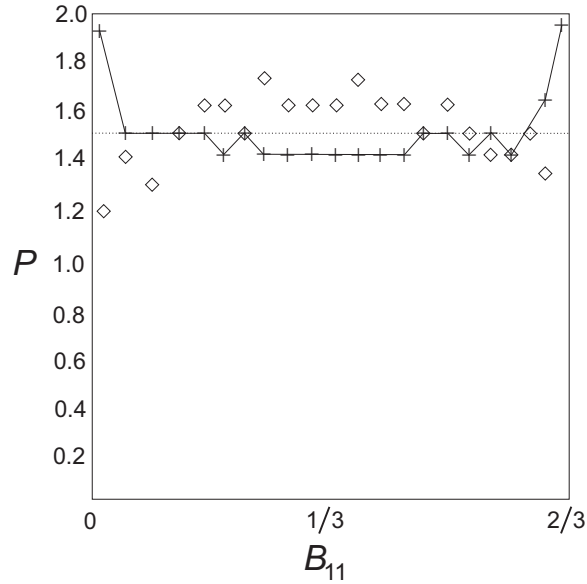}
\caption[]
{Probability density along the line $B_{12} = \frac{1}{3}$
of Fig.~\ref{figW1} obtained from $5\times 10^3$ events  for two
initial measures: {\bf (a)} the critical parameter $s=s^{\star}$\,
(marked by $+$ and decorated by a solid line to guide the eye) and {\bf (b)} the flat measure $s=1$ (marked by
$\Diamond$)\,.} \label{figW2}
\end{center}
\end{figure}

This feature is further visible in Fig~\ref{figW2}\,, which shows a
comparison of the results obtained for two different initial
measures on a one--dimensional cross section of Fig.~\ref{figW1}\,.
Although the measure obtained for the critical parameter $s^{\star}$
is indeed uniform in the vicinity of the center, namely around
$B_{11}=1/3$\,, the measure induced by random stochastic matrices
with the flat measure, $s=1$, displays similar properties. Since for
larger matrix size $N$ the value of the optimal  parameter
$s^{\star}$ tends to unity as $1-1/N^2$, it seems reasonable to
generate random bistochastic matrices of a larger size initiating
the iterative Algorithm (1) with random stochastic matrices distributed
according to the uniform measure, (i.e. each column is generated
independently according to the Dirichlet distribution with $s=1$).

\section{Estimation of the volume of the Birkhoff polytope}\label{sec5}

The set ${\cal B}_N$ of bistochastic matrices of size $N$ forms a
convex polytope in $\IR^{{\pt{N-1}}^2}$. Its volume with
respect to the Euclidean measure is known for
$N=2\,,...,\,10$~\cite{Cha99:1,Bec03:1}. The concrete numbers depend
on the normalization chosen. For instance, in the simplest case the
set ${\cal B}_2$ forms an interval $d\in [0,1]$, any point of which
corresponds to the bistochastic matrix,  $B(d)=\left(
                                                 \begin{smallmatrix}
                                                   d & 1-d \\
                                                   1-d & d \\
                                                 \end{smallmatrix}
                                               \right)$. If
the range of the single, independent element is concerned, the
\emph{relative volume} of the polytope
 reads $\nu\pt{ {\cal B}_2}=1$.
On the other hand, if we regard this set as an interval in
$\IR^4$, its length is equal to the \emph{volume} of the Birkhoff
polytope, $\mathrm{Vol}\pt{{\cal B}_2}=\sqrt{4}=2$. In general, both
definitions of the volumes are related by \cite{Can07:1}
\begin{equation}
 \mathrm{Vol}\pt{{\cal B}_N} \ = \ N^{N-1}\; \nu\pt{{\cal B}_N} \ .
 \label{relative}
\end{equation}
In Section~\ref{distB} we derived formula~\eqref{eq66}\,, giving the
probability distribution $P^{\star}_N$ at the center $B_N^{\star}$
of the Birkhoff polytope induced by the Dirichlet measure on the
space of input stochastic matrices.
If all Dirichlet parameters are equal to $s_i=s$ for $i=1,\dots N$ then formula~\eqref{eq66} simplifies to
\begin{equation}
 P^{\star}_N\pt{s}=
 \frac{\Gamma\pt{Ns}^{2N}}{\Gamma\big(N^2s\big)^{\phantom{N^2}}\!\!\!\Gamma\big(s\big)^{N^2}}\;N^{N^2-1}\,,
\label{Pstar}
\end{equation}
\noindent
Making use of the Stirling expansion
\begin{equation}
\Gamma\pt{x} \approx  \sqrt{2 \pi} \; x^{\,x-1/2} \; e^{-x} \;
\pq{1+\frac{1}{12x} +
 {\cal O} \pt{ \frac{1}{x^2}} } \ ,
 \label{stirling}
\end{equation}
and plugging it into eq.~\eqref{Pstar}
we obtain an approximation valid for a large matrix size $N$,
\begin{equation}
 P_N^{\star}  \ \approx\
N^{N^2-N}\;  \pt{2 \pi}^{\,N-1/2}\; s^{\,sN^2-N+1/2} \; \pq{\Gamma
(s)}^{-N^2}\;  \exp\pg{-sN^2+\frac{1}{6s} + {\cal O} \pt{
\frac{1}{N}} } \,.
 \label{eq66r}
\end{equation}

For $s=s^{\star}=1-1/N^2+{\cal O} (1/N^4)$ this distribution is flat
in the vicinity of the center $B_N^{\star}$ -- compare
eq.~\eqref{eq91}\,. Assuming it is close to uniform in the entire
Birkhoff polytope, we obtain an approximation of its relative
volume, $\nu \pt{{\cal B}_N}\approx 1/P_N^{\star}$. Substituting
$s^{\star}$ into~\eqref{eq66r} we arrive at
\begin{equation}
\nu ( {\cal B}_N) \approx N^{N-N^2}\;  (2 \pi)^{1/2-N} \;
\exp\pg{N^2 +C + {\cal O} \pt{ \frac{1}{N}} } \,.
 \label{eq66s}
\end{equation}
Making use of the expansion $\Gamma(1+x)=1-\gamma x+{\cal O}(x^2)$
we can express the value of $C$ by the Euler gamma constant
$\gamma\approx 0.\,577\;215\;665\ldots$ The result is
$C=\gamma-1/6\approx 0.\,410\;548\;998\ldots$.

Interestingly, the above approximation is identical, up to a value
of this constant, with the recent result of Canfield and
Mackay~\cite{Can07:1}. Making use of the relation~\eqref{relative}
we see that their asymptotic formula for the volume ${\rm vol} (
{\cal B}_N)$ of the Birkhoff polytope is consistent with eq. ~(\ref{eq66s})
for $C=1/3$\,. This fact provides a strong argument that the
distribution generated by the Dirichlet measure with
$s=s^{\star}$\,, is close (but not equal) to the uniform
distribution inside the Birkhoff polytope. Furthermore, the
initially flat distribution of the stochastic matrices, obtained for
$s=1$, leads to yet another reasonable approximation for the
relative volume of ${\cal B}_N$\,, equivalent to~\eqref{eq66s} with
$C=-1/6$\,.

\section{Concluding Remarks}

In this paper we introduced several ensembles of random stochastic matrices.
Each of them can be considered as an ensemble of initial points
used as  input data for the Sinkhorn Algorithm, which
generates  bistochastic matrices.
Thus any probability measure $W[M]$ in the set of stochastic matrices induces
a certain probability measure $P[B]$ in the set of bistochastic matrices.

Let us emphasize that the iterative procedure of Sinkhorn \cite{S64:1}
applied in this work,
covers the entire set of bistochastic matrices.
This is not the case for the ensemble of
\emph{unistochastic matrices}, which are obtained from a unitary
matrix by squaring moduli of its elements. Due to unitarity of $U$
the matrix $B_{ij}={\abs{U_{ij}}}^2$ is bistochastic, and the Haar
measure on $U(N)$ induces
a certain measure inside the Birkhoff polytope~\cite{Zyk03:1}.
However, for $N\geq 3$, this measure does not cover the entire
Birkhoff polytope since in this case there exist bistochastic
matrices which are not unistochastic~\cite{Mar79:1,Zyk03:1}.

In the general case of arbitrary $N$ we derive an
integral expression
representing the probability distribution inside the
$(N-1)^2$--dimensional Birkhoff polytope ${\cal B}_N$ of bistochastic matrices.
In the simplest case of $N=2$ it is straightforward to
obtain explicit formulae for the probability distribution
in the set of bistochastic matrices induced by the ensemble
of stochastic matrices, in which both columns are independent.
Furthermore,  we find that to
generate the uniform (flat) measure, $P[B]=\text{const}$\,, one needs
to start with random stochastic matrices of size $2$ distributed
according to eq.~\eqref{eq27}\,, for which both columns are
correlated.

For an arbitrary $N$ the integral form for the probability distribution
 can be explicitly worked out for a
particular point --- the flat, van der Waerden matrix~\eqref{eq65bis}
located at the center of the Birkhoff polytope.  In this case
we obtain an explicit formula for the probability distribution at
this point as a function of the parameters $\{s_i \mid 1\leq i\leq
N\}$ defining the Dirichlet distribution for each column of the
initially random stochastic matrix.
Expanding the probability density in the vicinity of $B^{\star}_N$ we
find the condition for the optimal parameters $s_i=s^{\star}$\,, for
which the density $P[B]$ is flat in this region. Discrepancy of the
measure constructed in this way from the uniform distribution is
numerically analyzed in the case $N=3$.

This measure is symmetric with respect to permutations of rows and
columns of the matrix and for large $N$ it tends to the uniform
measure in the set of bistochastic matrices. For large $N$ the
optimal Dirichlet  parameter $s^{\star}$ tends to unity as
$1-1/N^2$. Thus we may suggest a simplified procedure of taking the
initial stochastic matrices according to the flat measure, $(s=1)$.
Each column of such a random stochastic matrix is drawn
independently and it consists of $N$ numbers distributed uniformly in
the simplex $\Delta_{N-1}$. With an initial matrix constructed in
this way we are going to run Algorithm (1). Such a procedure is
shown to work fine already for $N=3$. We tend to believe that this
scheme of generating random bistochastic matrices could be useful
for several applications in mathematics, statistics and physics.

Assuming that a given probability measure in a compact set is flat, the value of
the probability density $P$ at an arbitrary point $x$ gives us an
information about the Euclidean volume of this set, $V=1/P(x)$\,. We
were pleased to find that the optimal algorithm for generating
random bistochastic matrices is characterized by an inverse
probability $1/P_s[B^{\star}_N]$ at the center $B^{\star}_N$ of the
polytope which displays the same dependence on the dimension $N$ as
the volume of the Birkhoff polytope, Vol$({\cal B}_N)$, derived
in~\cite{Can07:1}.

Although in this paper we analyzed dynamics in the classical
probability simplex, the main idea of the algorithm may be
generalized for the quantum dynamics. In such a case a stochastic
matrix corresponds to a stochastic map (so called quantum
operation), which sends the set of quantum states (Hermitean,
positive matrices of trace one) into itself~\cite{Ben79:1}.  A
quantum stochastic map is called \emph{bistochastic}, if it
preserves the maximally mixed state, ${\Id}/N$. To generate random
bistochastic maps one can use an analogous technique of alternating
projection onto the subspaces in which a given map or its dual is
stochastic. Such an algorithm suitable for the quantum problem,
was  proposed independently by Audenaert and Scheel \cite{Aud07:1}.
First results concerning various measures induced into the
set of quantum stochastic maps are presented in~\cite{Som07:1}.

\section*{Acknowledgements}
 {The authors gratefully acknowledges financial support provided by
 the \textsf{EU} Marie Curie Host Fellowships for Transfer of Knowledge
 Project \co{COCOS} (contract number \textsf{MTKD}--\textsf{CT}--2004--517186)
 and the \co{SFB}/Transregio--12 project financed by
 \co{DFG} and the Polish Ministry of Science
under the grant number 1~P03B~042~26\,.}

\section*{Appendix A}
\label{app1}

In this appendix we demonstrate that
the Algorithm (2) suitable for analytical calculations is
 equivalent with the Sinkhorn Algorithm (1).

To apply the former Algorithm (2) one takes some
initial matrix $M\in End\pq{\IR_+^{N}}$ and makes it bistochastic by
means of left-- and right--multiplication by two matrices $D^{L}$\,,
and $D^{R}$\,. The latter are limits of convergent
sequences of diagonal matrices $D^{L}=\lim_nD^{L}_n$\,and
$D^{R}=\lim_nD^{R}_n$ and the finally $B=D^{L}MD^{R}$\,.

In a similar way, Algorithm (1) performs the same task of
transforming the initially stochastic matrix $M\in End\pq{\IR_+^{N}}$ into a bistochastic matrix
$B$ by alternating rows-- and columns--normalization (\co{R} and
\co{C}, for short), which in turn is the same of left--\,,
respectively right--multiplication by diagonal matrices. Once a matrix $M=\{m_{pq}\geq 0\}$ is
given to renormalize the $p^{\text{th}}$ row means to divide each of
its elements by the factor $\sum_q m_{pq}$\,,
\begin{subequations}\label{eq47}
 \begin{align}
 m_{pq}\longmapsto m^{\prime}_{pq}=\hat{L}_p^{\phantom{\prime}}\,m^{\phantom{\prime}}_{pq}
 \quad,\quad&\text{with}\qquad\frac{1}{\hat{L}_p}=\sum_q m_{pq}\,.
 \label{eq47a}
 \intertext
 {Analogously, to renormalize the $q^{\text{th}}$ column means to divide each of its elements by the factor $\sum_p m_{pq}$\,, }
 m_{pq}\longmapsto m^{\prime}_{pq}=m^{\phantom{\prime}}_{pq}\,\hat{R}_q^{\phantom{\prime}}
 \quad,\quad&\text{with}\qquad\frac{1}{\hat{R}_q}=\sum_p m_{pq}
 \label{eq47b}
 \,.
\end{align}
\end{subequations}
Let us now run the Algorithm~\ref{alg1}, taking as an input a generic
$M^{(0)}=\{m_{pq}^{(0)}\geq 0\}$\,, and set \co{1CRCRCR\ldots} to be
the row--column renormalization sequence, where the first symbol
\co{1} denotes the  dummy operation
\begin{subequations}\label{eq48}
 \begin{align}
 \frac{1}{\hat{L}_p^{(0)}}=1
 \qquad & \blacklozenge\qquad
 \nullla{m}_{pq}^{(0)}\longmapsto \nulllb{m}_{pq}^{(0)}=
 \hat{L}_p^{(0)}\,\nullla{m}_{pq}^{(0)}\label{eq48a}\,.
 \intertext{Now we start with equations~\eqref{eq47b}}
 \frac{1}{\hat{R}_q^{(0)}}=
 \sum_p \nulllb{m}_{pq}^{(0)}=
 \sum_p \hat{L}_p^{(0)}\,\nullla{m}^{(0)}_{pq}
 \qquad & \blacklozenge\qquad
 \nulllb{m}_{pq}^{(0)}\longmapsto \nullla{m}_{pq}^{(1)}=
 \nulllb{m}_{pq}^{(0)}\,\hat{R}_q^{(0)}=
 \hat{L}_p^{(0)}\,\nullla{m}_{pq}^{(0)}\,\hat{R}_q^{(0)}\,,
 \intertext{followed by~\eqref{eq47a}}
 \frac{1}{\hat{L}_p^{(1)}}=
 \sum_q \nullla{m}^{(1)}_{pq}=
 \sum_q \hat{L}_p^{(0)}\,\nullla{m}_{pq}^{(0)}\,\hat{R}_q^{(0)}
 \qquad & \blacklozenge\qquad
 \nullla{m}_{pq}^{(1)}\longmapsto \nulllb{m}_{pq}^{(1)}=
 \hat{L}_p^{(1)}\,\nullla{m}_{pq}^{(1)}
 \label{eq48c}\,.
 \intertext{The next two steps are}
 \frac{1}{\hat{R}_q^{(1)}}=
 \sum_p \nulllb{m}_{pq}^{(1)}=
 \sum_p \hat{L}_p^{(1)}\,\nullla{m}^{(1)}_{pq}
 \qquad & \blacklozenge\qquad
 \nulllb{m}_{pq}^{(1)}\longmapsto \nullla{m}_{pq}^{(2)}=
 \nulllb{m}_{pq}^{(1)}\,\hat{R}_q^{(1)}=
 \hat{L}_p^{(1)}\,\nullla{m}_{pq}^{(1)}\,\hat{R}_q^{(1)}=
 \hat{L}_p^{(1)}\,\hat{L}_p^{(0)}\,\nullla{m}_{pq}^{(0)}\,\hat{R}_q^{(0)}\,\hat{R}_q^{(1)}\label{eq48d}
 \intertext{and}
 \frac{1}{\hat{L}_p^{(2)}}=
 \sum_q \nullla{m}^{(2)}_{pq}=
 \sum_q
 \quad
 \hat{L}_p^{(1)}\,\hat{L}_p^{(0)}\,&\nullla{m}_{pq}^{(0)}\,\hat{R}_q^{(0)}\,\hat{R}_q^{(1)}
 \qquad\blacklozenge\qquad\cdots
 \label{eq48e}
\end{align}
\end{subequations}
so that the iteration procedure can be written as
\begin{equation}
\begin{cases}
 \quad\displaystyle\frac{\displaystyle 1}{\displaystyle\hat{L}_p^{(n)}\hat{L}_p^{(n-1)}\cdots\hat{L}_p^{(1)}\hat{L}_p^{(0)}} & =
 \displaystyle\sum_q \ \ \nullla{m}_{pq}^{(0)}\ \ \hat{R}_p^{(0)}\hat{R}_q^{(1)}\cdots\hat{R}_q^{(n-1)}\\
 \quad\displaystyle\frac{\displaystyle 1}{\displaystyle\hat{R}_p^{(n)}\hat{R}_q^{(n-1)}\cdots\hat{R}_q^{(1)}\hat{R}_q^{(0)}} & =
 \displaystyle\sum_p\ \ \hat{L}_p^{(n)}\hat{L}_p^{(n-1)}\cdots\hat{L}_p^{(1)}\hat{L}_p^{(0)} \ \ \nullla{m}_{pq}^{(0)}\label{eq49}\,.
\end{cases}
\end{equation}
The latter form can be rewritten more compactly,
\begin{equation}
 \quad\displaystyle\frac{\displaystyle 1}{\displaystyle\check{L}_p^{(n)}}  =
 \displaystyle\sum_q \ \ \nullla{m}_{pq}^{(0)}\ \ \frac{\displaystyle 1}{\displaystyle\sum_s\ \ \check{L}_s^{(n-1)} \ \ \nullla{m}_{sq}^{(0)}}
 \label{eq50}\,,
\end{equation}
 where we introduced new variables

\begin{equation}
 \check{L}_s^{(n)}\coleq\prod_{\ell=1}^n
 \hat{L}_s^{(\ell)}
 \qquad\text{and}\qquad
 \check{R}_s^{(n)}\coleq\prod_{\ell=1}^n
 \hat{R}_s^{(\ell)}
 \label{eq51_2}\,.
\end{equation}
 Equation~\eqref{eq49} is formally equivalent
 to~\eqref{eq45}, the only difference being in the number of
 component of $L$ vectors, respectively $\check{L}$, that are
 processed: in Algorithm~(1) one iterates all $\check{L}_s^{(n)}$, whereas in
 Algorithm~(2) the element $L_N^{(n)}$ is fixed to unity in each step.
We know that the solution of the limit equation for $L^{(n)}$ is not unique. But the only
non-uniqueness is due to multiplication by a fixed factor $\eta>0$.

\section*{Appendix B}
\label{app2}

In this appendix we present the basic steps allowing one
to derive the central result of this work -
the second order expansion (\ref{eq89}) around the center of the
Birkhoff polytope of the probability distribution generated by
Dirichlet random stochastic  matrices.

Since $\widetilde{P}_{\bs{s}}\big[\{B_{ij}\}\big]\geq
0$\,, it is convenient to expand $\ln
\widetilde{P}_{\bs{s}}\big[\{B_{ij}\}\big]$\,. We denote the sum of the Dirichlet parameters for each column by
$\sigma=\sum_{j=1}^N s_j$ and start with the following integral
\begin{multline}
 Q_{\bs{s}}\big[\{B_{ij}\}\big]\coleq\int_0^\infty\!\!\!\!\!\cdots\!\!\int_0^\infty\pt{\prod_{r=1}^N\ud
 \alpha_r\;\alpha_r^{\,\sigma-1}}\;
 \prod_{w=1}^N \;\frac{1}{\pq{\sum_{h}\alpha_h \,\frac{1}{N} + \sum_{h}\alpha_h \,\delta B_{hw}}^{\,Ns_w}}\;\,
 \delta\pt{\alpha_N-1}=\\
 =\int_0^\infty\!\!\!\!\!\cdots\!\!\int_0^\infty\pt{\prod_{r=1}^N\ud
 \alpha_r\;\alpha_r^{\,\sigma-1}}\;\frac{1}{\pt{\sum_{h}\alpha_h \,\frac{1}{N}}^{\,N\sigma}}
 \exp\pq{-\sum_{w=1}^N {\,Ns_w}\;\ln\pt{1 + \frac{\sum_{h}\alpha_h \,\delta B_{hw}}{\sum_{h}\alpha_h \,\frac{1}{N}}}}\;\,
 \delta\pt{\alpha_N-1}\label{eq73}\,.
\end{multline}
Expanding the function \ $\ln\pt{1+x}=x-\frac{x^2}{2}+{\cal
O}\pt{x^3}$\,,
\begin{equation}
 \exp\pq{-\sum_{w=1}^N {\,Ns_w}\;\ln\pt{1 + \frac{\sum_{h}\alpha_h
 \,\delta B_{hw}}{\sum_{h}\alpha_h \,\frac{1}{N}}}}=
 \exp\pg{-\sum_{w=1}^N {\,Ns_w}\;\pq{
 \frac{\sum_{h}\alpha_h \,\delta B_{hw}}{\sum_{h}\alpha_h \,\frac{1}{N}}
 -\frac{1}{2}\pt{\frac{\sum_{h}\alpha_h \,\delta B_{hw}}{\sum_{h}\alpha_h \,\frac{1}{N}}}^2
 +
 {\cal O}\pt{\pt{\delta B}^3}}}\,,
\end{equation}
and then $\mathrm{e}^{-x}\approx 1-x+x^2/2$ we get
\begin{multline}
 Q_{\bs{s}}\big[\{B_{ij}\}\big]
 =N^{\,N\sigma}\int_0^\infty\!\!\!\!\!\cdots\!\!\int_0^\infty\pt{\prod_{r=1}^N\ud
 \alpha_r\;\alpha_r^{\,\sigma-1}}\;\frac{\delta\pt{\alpha_N-1}}{\pt{\sum_{h}\alpha_h}^{\,N\sigma}}
 \left\{1-
 N^2\; \frac{\sum_w {\,s_w}\sum_{h}\alpha_h \,\delta B_{hw}}{\sum_{h}\alpha_h }
 +\right.\\
 \left.
 +\frac{N^3}{2}\sum_w {\,s_w}\pt{\frac{\sum_{h}\alpha_h \,\delta B_{hw}}{\sum_{h}\alpha_h }}^2
 +\frac{N^4}{2}\pt{
 \frac{\sum_w {\,s_w}\sum_{h}\alpha_h \,\delta B_{hw}}{\sum_{h}\alpha_h
 }}^2
 +
 {\cal O}\pt{\pt{\delta B}^3}
 \right\}\,.
 \label{eq74}
\end{multline}
Thus we have to integrate the following expression for an arbitrary vector of parameters $\vartheta_w\geqslant 0$
\begin{multline}
 I\coleq\int_0^\infty\!\!\!\!\!\cdots\!\!\int_0^\infty\pt{\prod_{r=1}^N\ud
 \alpha_r\;\alpha_r^{\,\sigma-1}}\;\frac{\delta\pt{\alpha_N-1}}{\pt{\sum_{h}\alpha_h}^{\,N\sigma}}
 \;\prod_{w=1}^N \;\pt{\frac{\alpha_w }{\sum_{h}\alpha_h }}^{\vartheta_w}=\\
 =\frac{1}{\Gamma\pt{\sigma N+m}}\int_0^\infty\!\!\!\!\!\cdots\!\!\int_0^\infty\pt{\prod_{r=1}^N\ud
 \alpha_r\;\alpha_r^{\,\sigma-1}} \;\alpha_1^{\,\vartheta_1}\,\alpha_2^{\,\vartheta_2}
 \cdots\alpha_{N-1}^{\,\vartheta_{N-1}}\;\delta\pt{\alpha_N-1}
 \int_0^{\infty}\ud
 t\;\mathrm{e}^{-t\sum_{h}\alpha_h}\;t^{\,N\sigma+m-1}\,,\label{eq74bis}
\end{multline}
Here $m\coleq
\vartheta_1+\vartheta_2+\cdots+\vartheta_{N-1}+\vartheta_N$\,, so the integral reads
\begin{equation}
I =\frac{\Gamma\pt{\sigma+\vartheta_1}\,
 \Gamma\pt{\sigma+\vartheta_2}\ldots\Gamma\pt{\sigma+\vartheta_{N-1}}\,\Gamma\pt{\sigma+\vartheta_N}}{\Gamma\pt{\sigma N+m}}=
 \frac{\Gamma\pt{\sigma}^N}{\Gamma\pt{\sigma N}}\left\langle \prod_{w=1}^N \;\pt{\frac{\alpha_w }{\sum_{s}\alpha_s
 }}^{\vartheta_w}\right\rangle\,,
\end{equation}
with
\begin{equation}
 \left\langle \prod_{w=1}^N \;\pt{\frac{\alpha_w }{\sum_{h}\alpha_h
 }}^{\vartheta_w}\right\rangle\coleq
 \frac{\Gamma\pt{\sigma N}}{\Gamma\pt{\sigma N+\sum_i \vartheta_i}}
 \;\prod_{j=1}^N
 \;\frac{\Gamma\pt{\sigma+\vartheta_j}}{\Gamma\pt{\sigma}}\label{eq75}\,\cdot
\end{equation}
This expression, completely symmetric in all variables $\alpha_1$\,,
$\alpha_2$\,\ldots\,$\alpha_{N-1}$\,, $\alpha_N$\, allows us to calculate the expansion of the integral~\eqref{eq74}\,:
\begin{multline}
 Q_{\bs{s}}\big[\{B_{ij}\}\big]
 =N^{\,N\sigma}\;\frac{\Gamma\pt{\sigma}^N}{\Gamma\pt{\sigma N}}
 \left\{1-
 N^2\; \left\langle\frac{\sum_w {\,s_w}\sum_{h}\alpha_h \,\delta B_{hw}}{\sum_{h}\alpha_h }\right\rangle
 +\right.\\
 \left.
 +\frac{N^3}{2}\sum_w {\,s_w}\left\langle\pt{\frac{\sum_{h}\alpha_h \,\delta B_{hw}}{\sum_{h}\alpha_h }}^2\right\rangle
 +\frac{N^4}{2}\left\langle\pt{
 \frac{\sum_w {\,s_w}\sum_{h}\alpha_h \,\delta B_{hw}}{\sum_{h}\alpha_h
 }}^2\right\rangle
 +
 {\cal O}\pt{\pt{\delta B}^3}
 \right\}\,.
 \label{eq76}
\end{multline}
Therefore we need
\begin{equation}
\begin{cases}
 \quad\rule[-17pt]{0pt}{7pt}\displaystyle\left\langle\frac{\alpha_u }{\sum_{h}\alpha_h }\right\rangle & =
 \;\displaystyle\frac{\Gamma\pt{\sigma+1}}{\Gamma\pt{\sigma}}
 \cdot\frac{\Gamma\pt{\sigma N}}{\Gamma\pt{\sigma N+1}}=\frac{\sigma}{\sigma N}=\frac{1}{N}
 \\
 \quad\rule[-17pt]{0pt}{7pt}\displaystyle\left\langle\frac{\alpha_u^2 }{{}^{\phantom{2}}\pt{\sum_{h}\alpha_h}^2}\right\rangle & =
 \;\displaystyle\frac{\Gamma\pt{\sigma+2}}{\Gamma\pt{\sigma}}
 \cdot\frac{\Gamma\pt{\sigma N}}{\Gamma\pt{\sigma N+2}}=\frac{\sigma\pt{\sigma+1}}{\sigma N\pt{\sigma N+1}}=\frac{\sigma+1}{N\pt{\sigma N+1}}
 \\
 \quad\rule[-7pt]{0pt}{7pt}\displaystyle\left\langle\frac{\alpha_u \alpha_v }{{}^{\phantom{2}}\pt{\sum_{h}\alpha_h}^2}\right\rangle & =
 \;\displaystyle\frac{\Gamma\pt{\sigma+1}}{\Gamma\pt{\sigma}}\frac{\Gamma\pt{\sigma+1}}{\Gamma\pt{\sigma}}
 \cdot\frac{\Gamma\pt{\sigma N}}{\Gamma\pt{\sigma N+2}}=\frac{\sigma\cdot \sigma}{\sigma N\pt{\sigma N+1}}=\frac{\sigma}{N\pt{\sigma N+1}}
\end{cases}\label{eq77}\,\cdot
\end{equation}
Thus the second term in~\eqref{eq76} is
\begin{align}
 \left\langle\frac{\sum_w {\,s_w}\sum_{h}\alpha_h \,\delta
 B_{hw}}{\sum_{i}\alpha_i
 }\right\rangle
 &=
 \frac{1}{N}\sum_{hw}\,\delta B_{hw}\,s_w
 =0\quad,\quad\text{because of~\eqref{eq72}}\,\cdot\label{eq78}
 \end{align}

\noindent For the third term we have
\begin{align}
 J\coleq\sum_w {\,s_w}\left\langle\pt{\frac{\sum_{h}\alpha_h \,\delta B_{hw}}{\sum_{i}\alpha_i }}^2
 \right\rangle
=
 \sum_w {\,s_w}\sum_{h^{\phantom{\prime}}}\sum_{h^{\prime}\neq h^{\phantom{\prime}}}\left\langle\frac{\alpha_{h^{\phantom{\prime}}}
 \alpha_{h^{\prime}}}{{}^{\phantom{2}}{\pt{\sum_{h}\alpha_h}^2}}
 \right\rangle \,\delta B_{h^{\phantom{\prime}}\!\!w}\:\delta B_{h^{\prime}\!w}
 +
 \sum_w
 {\,s_w}\sum_{h^{\phantom{\prime}}}\left\langle\frac{\alpha_{h}^2}{{}^{\phantom{2}}\pt{\sum_{h}\alpha_h}^2}
 \right\rangle \,{\pt{\delta B_{hw}}}^2\nonumber\\
=
 \sum_w \frac{\sigma{\,s_w}}{N\pt{\sigma N+1}}\sum_{h^{\phantom{\prime}}}\:\delta B_{h^{\phantom{\prime}}\!\!w}\sum_{h^{\prime}\neq h^{\phantom{\prime}}}
 \:\delta B_{h^{\prime}\!w}
 +
 \sum_w
 \frac{\pt{\sigma+1}{\,s_w}}{N\pt{\sigma N+1}}\sum_{h^{\phantom{\prime}}} \,{\pt{\delta B_{hw}}}^2\nonumber
 \intertext{and using from~\eqref{eq72} the relation $\sum_{h^{\prime}\neq h^{\phantom{\prime}}}
 \:\delta B_{h^{\prime}\!w}=-\delta B_{h^{\phantom{\prime}}\!\!w}$\, we get}
J =
 \sum_w {\,s_w}\sum_{h^{\phantom{\prime}}} \,{\pt{\delta
 B_{hw}}}^2\:\pt{\frac{\pt{\sigma+1}}{N\pt{\sigma N+1}}-\frac{\sigma}{N\pt{\sigma N+1}}}=\:\frac{1}{N\pt{\sigma N+1}}
 \:\sum_w {\,s_w}\sum_{h^{\phantom{\prime}}} \,{\pt{\delta
 B_{hw}}}^2\,\cdot
 \label{eq81}
 \intertext{For the fourth term we obtain}
 \left\langle\pt{
 \frac{\sum_{h}\alpha_h \,\sum_w {\,s_w}\,\delta B_{hw}}{\sum_{h}\alpha_h
 }}^2\right\rangle
 =
 \frac{1}{N\pt{\sigma N+1}}\sum_{h^{\phantom{\prime}}} \,{\pt{\sum_w
 {\,s_w}\,\delta B_{hw}}}^2\,\cdot
 \label{eq82}
\end{align}
Finally, expression~\eqref{eq76} yields:
\begin{equation}
 Q_{\bs{s}}\big[\{B_{ij}\}\big]
 =N^{\,N\sigma}\;\frac{\Gamma\pt{\sigma}^N}{\Gamma\pt{\sigma N}}
 \left\{1
 +\frac{N^2}{2\pt{\sigma N+1}}\:\sum_{h^{\phantom{\prime}}\!\!w} {\,s_w} \,{\pt{\delta
 B_{hw}}}^2
 +\frac{N^3}{2\pt{\sigma N+1}}\sum_{h^{\phantom{\prime}}} \,{\pt{\sum_w
 {\,s_w}\,\delta B_{hw}}}^2
 +
 {\cal O}\pt{\pt{\delta B}^3}
 \right\}\,\cdot
 \label{eq83}
\end{equation}
In principle we are able to calculate all higher terms. There are
two other terms to be expanded: $\prod_{p,q=1}^N
 {B_{pq}}^{s_q-1}$ and $\det{\pq{\Id-BB^{\text{\co{T}}}}}_{N-1}$\,.
 For the latter we have
\begin{align}
 \det{\pq{\Id-BB^{\text{\co{T}}}}}_{N-1}
 =
 \det{\pq{\delta_{ik}-\sum_{j=1}^{N} \pt{\frac{1}{N} +  \delta B_{ij}}\,\pt{\frac{1}{N} + \delta B_{kj}}
 }}_{N-1}
 =\exp\pg{
 \ln\det{\pq{D_{ik}
 -\sum_{j=1}^{N}\delta B_{ij}\,\delta B_{kj}
 }}_{N-1}} \label{eq84}\,\cdot
\end{align}
In the last line we made use of equation~\eqref{eq72} and we
introduced the $\pt{N-1}\times \pt{N-1}$
\emph{circulant}~\cite{Hof98:1} matrix $D_{ik}\coleq \delta_{ik}
 -{1}/{N}$\,. As it can be verified by direct matrix multiplication,
 the inverse of $D$ reads $D_{ik}^{-1}=\delta_{ik}^{\phantom{-1}}\!\!+1$. Hence,
 factorizing the determinant of the product in the product of determinants,
 it follows from~\eqref{eq84}
\begin{align}
 \det{\pq{\Id-BB^{\text{\co{T}}}}}_{N-1}
 =\det{\Big[D\Big]}_{N-1}\times\det{\pq{\delta_{i\ell}
 -\sum_{j=1}^{N-1}\sum_{k=1}^N \pt{\delta_{ij}+1}\delta B_{jk}\,\delta B_{\ell k}
 }}_{N-1}\label{eq85}\,\cdot
 \intertext
 {Observe that the index $j$ labels the $\pt{N-1}$ columns of the matrix $\pq{D^{-1}}_{N-1}$\,,
 whereas $k$ runs from $1$ to $N$\,, since we are going to consider
  $\pq{\pt{\delta B}\,\pt{\delta B}^{\text{\co{T}}}}_{N-1}$ and not $\pq{\pt{\delta B}^{\text{\co{\phantom{T}}}}\!\!}_{N-1}\,
  \pq{\pt{\delta B}^{\text{\co{T}}}}_{N-1}$\,. Using the property of circulant matrices~\cite{Hof98:1}, we can determine the spectrum of
  \rule[0pt]{0pt}{10pt}$D$\,, consisting of a simple eigenvalue $1/N$ and another one equal to $1$\,,
  of multiplicity $\pt{N-2}$. Thus $\det{\Big[D\Big]}_{N-1}=1/N$
  and eq.~\eqref{eq72} and~\eqref{eq85} yield}
 \det{\pq{\Id-BB^{\text{\co{T}}}}}_{N-1} =\frac{1}{N}\times\det{\pq{\delta_{i\ell}
 -\sum_{k=1}^N \delta B_{ik}\,\delta B_{\ell k}
 +\sum_{k=1}^N \delta B_{Nk}\,\delta B_{\ell k}
 }}_{N-1}
 \,\cdot
 \label{eq86}
\end{align}
 From the identity $\det\pq{\exp\pt{A}}=\exp\pq{\tr\pt{A}}$\,, with the substitution
 $A\leftarrow\log\pt{\Id+X}$\,
 we get
\begin{align}
 \det\pt{\Id-X}&=\exp\pg{\tr\pq{\log\pt{\Id-X}}}=\exp\pg{\tr\pq{-X+{\cal
 O}\pt{X^2} }}\nonumber\\
 &=\exp\pg{-\tr\pt{X}+{\cal
 O}\pq{\tr\pt{X^2}} }=1-
 \tr\pt{X}+\frac{\pq{\tr\pt{X}}^2}{2}+{\cal O}\pq{\tr\pt{X^2}}
 \nonumber
\end{align}
so that, choosing for $X$ the $\pt{\delta B}$'s contributions in
equation~\eqref{eq86}\,, we get $\tr\pt{X}=\sum_{\ell,k=1}^N \delta
B_{\ell k}\,\delta B_{\ell k}$ and therefore
\begin{equation}
 \det{\pq{\Id-BB^{\text{\co{T}}}}}_{N-1} =\frac{1}{N}\,\left\{1
 -\:\sum_{\ell k}  \,{\pt{\delta
 B_{\ell k}}}^2
 +
 {\cal O}\pt{\pt{\delta B}^4}
 \right\}\,.
 \label{eq87}
\end{equation}
Finally we use the expansion
\begin{align}
 \prod_{p,q=1}^N {B_{pq}}^{s_q-1}&=\exp\pg{\sum_{p,q=1}^N \pt{s_q-1}\ln\pt{\frac{1}{N} +  \delta
 B_{pq}}}
 =\frac{1}{N^{N\pt{\sigma-N}}}\times\pg{
 \;1+\frac{N^2}{2}\sum_{p,q=1}^N{\pt{\delta B_{pq}}}^2-\frac{N^2}{2}\sum_{p,q=1}^N s_q\,
 {\pt{\delta B_{pq}}}^2 +{\cal O}\pt{\pt{N\,\delta
 B}^3}}\label{eq88}\,.
\end{align}
Now, substituting~\eqref{eq83}\,and~(\ref{eq87}--\ref{eq88})
into~\eqref{eq71}\,, we obtain the final formula for the resulting probability distribution around the center
of the Birkhoff polytope ${\cal B}_N$ given by~(\eqref{eq89}).

\section*{Appendix C}
\label{app3}

In this appendix we provide the third order expansion of the
probability distribution $P_{\bs s}(B)$ at $B=B_{\star}$. The
result obtained implies that it is not possible to find an ensemble
of stochastic matrices characterised by the Dirichlet distribution,
which induces a distribution flat up to the third order at the
center of the Birkhoff polytope. Furthermore, we provide an
estimation, that is how the asymmetry of the optimal
distribution around $B_{\star}$ changes with $N$.

For general $s$ the output distribution behaves like
 $\widetilde{P}_{s}\big[\{B_{ij}\}\big]\propto\exp\pt{\lambda \sum_{pq}\,{\pt{\delta
 B_{pq}}}^2}$ at the center, with
 \begin{equation}
 \lambda=\frac{N^2}{2}-1-\frac{N^4\,s^2}{2\pt{N^2\,s+1}}\,\cdot\label{eq92}
 \end{equation}
From now on, symbols like
$\widetilde{P}_{s}\,,\,P_s\,,\,V_s\,,\,W_s$ denote the
probability densities obtained from the input described by the string\\
$\bs{s}=\pg{\,s_1=s,s_2=s,\ldots,s_N=s\,}$ consisting of $N$
Dirichlet exponents equal.
Since $\frac{\ud \lambda}{\ud s}<0$\,, the distribution is Gaussian
for $s>s^{\star}$\,.

In order to study the deviations from the Gaussian distribution, we
now study the third order contribution to
$\widetilde{P}_{s}\big[\{B_{ij}\}\big]$ of eq.~\eqref{eq71}\,, in
the case $s_i=s$\,. Under the latter hypothesis, many
terms of the kind $\sum_q
 {\,s_q}\,\delta B_{pq}$ vanish for~\eqref{eq72}\, so such terms will be omitted.

The distribution~\eqref{eq71} can be factorized into a product of three factors:
\begin{itemize}
    \item $\prod_{p,q=1}^N {B_{pq}}^{s-1}$ gives a contribution
        \begin{equation}
        \frac{1}{N^{N\pt{\sigma-N}}}\times\pq{
        \;\frac{N^3}{3}\pt{s-1}\sum_{p,q=1}^N
        {\pt{\delta B_{pq}}}^3 +{\cal O}\pt{\pt{N\,\delta
        B}^4}}\label{eq93}\,;
        \end{equation}
    \item $\det{\pq{\Id-BB^{\text{\co{T}}}}}_{N-1}$ gives no
        $3^{\text{rd}}$ order contribution (just the overall factor $1/N$ already present
        in~\eqref{eq87})\,;
    \item the integral $Q_{s}\big[\{B_{ij}\}\big]$ of~\eqref{eq73} gives
        \begin{align}
        \Delta_3 Q_{s}\big[\{B_{ij}\}\big]
        &=N^{\,N\sigma}\int_0^\infty\!\!\!\!\!\cdots\!\!\int_0^\infty\pt{\prod_{r=1}^N\ud
        \alpha_r\;\alpha_r^{\,\sigma-1}}\;\frac{\delta\pt{\alpha_N-1}}{\pt{\sum_{s}\alpha_s}^{\,N\sigma}}
        \left\{
        -\frac{1}{3}Ns\sum_j \pt{\frac{N\sum_{i}\alpha_i \,\delta B_{ij}}{\sum_{k}\alpha_k
        }}^3
        + {\cal O}\pt{\pt{\delta B}^4} \right\}
        \nonumber\\
        &=N^{\,N\sigma}\,\frac{\Gamma\pt{\sigma}^N}{\Gamma\pt{\sigma N}}\pg{-\frac{N^4
        s}{3}\,\sum_j\,
        \left\langle\;\pt{\frac{\sum_{i}\alpha_i \,\delta B_{ij}}{\sum_{k}\alpha_k
        }}^3\right\rangle+ {\cal O}\pt{\pt{\delta B}^4}}\,,\label{eq94}
        \end{align}
        where we made use of the symbol $\left\langle\cdot\right\rangle$ introduced
        through eqs.~(\ref{eq74bis}--\ref{eq75})\,.
\end{itemize}
Using the same reasoning as in Appendix~B, including now the
new contributions~(\ref{eq93}--\ref{eq94}), we arrive at the
$3^{\text{rd}}$ order contribution for
$\widetilde{P}_{s}\big[\{B_{ij}\}\big]$\,,
        \begin{equation}
        \Delta_3 \widetilde{P}_{s}\big[\{B_{ij}\}\big]
        =P^{\star}_N\pg{\frac{N^3}{3}\pt{s-1}\sum_{p,q=1}^N
        {\pt{\delta B_{pq}}}^3-\frac{N^4
        s}{3}\,\sum_j\,
        \left\langle\;\pt{\frac{\sum_{i}\alpha_i \,\delta B_{ij}}{\sum_{k}\alpha_k
        }}^3\right\rangle}\,.\label{eq95}
        \end{equation}
The last term reads
        \begin{align}
        \left\langle\;\pt{\frac{\sum_{i}\alpha_i \,\delta B_{ij}}{\sum_{k}\alpha_k
        }}^3\right\rangle
        &=
        &\!\!\!\!\!\!\!\!\!\!=
        \left\langle\;\displaystyle\frac{
        \alpha_1 \,\alpha_2 \,\alpha_3}
        {\pt{\sum_{k}\alpha_k
        }^3}\right\rangle
        \displaystyle\sum_{\textstyle\newatop{\mu\neq\nu\,,\,\nu\neq\tau}{\tau\neq\mu}}
        \delta B_{\mu j}\,\delta B_{\nu j}\,\delta B_{\tau j}
        +
        3\,\left\langle\;\displaystyle\frac{
        \alpha_1^2 \,\alpha_2 }
        {\pt{\sum_{k}\alpha_k
        }^3}\right\rangle
        \displaystyle\sum_{\mu\neq\tau}
        \pt{\delta B_{\mu j}}^2\,\delta B_{\tau j}
        +
        \left\langle\;\frac{
        \alpha_1^3  }
        {\pt{\sum_{k}\alpha_k
        }^3}\right\rangle
        \displaystyle\sum_{\mu}\pt{\delta B_{\mu j}}^3
        \,.\label{eq96}
        \end{align}
It follows from~\eqref{eq72}\,, that
        $\displaystyle\sum_{\tau} \delta B_{\tau j}=0=
        \displaystyle\sum_{\tau\neq\mu}\delta B_{\tau j}+\delta B_{\mu
        j}$\,, so $\sum_{\tau\neq\mu}\delta B_{\tau j}=-\delta B_{\mu
        j}$\, Multiplying this equality by ${\pt{\delta B_{\mu j}}^2}$ and summing over $\mu$ one gets
        \begin{subequations}\label{eq97}
        \begin{align}
        \displaystyle\sum_{\mu\neq\tau}
        \pt{\delta B_{\mu j}}^2\,\delta B_{\tau j}&=-\displaystyle\sum_{\mu}
        \pt{\delta B_{\mu j}}^3\,.\label{eq97a}
        \intertext{Similarly}
        \pt{\displaystyle\sum_{\mu} \delta B_{\mu j}}^3=0=
        \displaystyle\sum_{\mu\,,\,\nu\,,\,\tau}\delta B_{\mu j}\,\delta B_{\nu j}\,\delta B_{\tau j}&=
        \displaystyle\sum_{\textstyle\newatop{\mu\neq\nu\,,\,\nu\neq\tau}{\tau\neq\mu}}
        \delta B_{\mu j}\,\delta B_{\nu j}\,\delta B_{\tau j}
        +
        3\,
        \displaystyle\sum_{\mu\neq\tau}
        \pt{\delta B_{\mu j}}^2\,\delta B_{\tau j}
        +
        \displaystyle\sum_{\mu}\pt{\delta B_{\mu j}}^3\nonumber
        \intertext{and using~\eqref{eq97a} we arrive at}
        \displaystyle\sum_{\textstyle\newatop{\mu\neq\nu\,,\,\nu\neq\tau}{\tau\neq\mu}}
        \delta B_{\mu j}\,\delta B_{\nu j}\,\delta B_{\tau j}
        &=2
        \displaystyle\sum_{\mu}\pt{\delta B_{\mu j}}^3\,.
        \label{eq97b}
        \end{align}
        \end{subequations}
        Substituting eqs.~\eqref{eq97} into~\eqref{eq96} one obtains
        \begin{align}
        \left\langle\;\pt{\frac{\sum_{i}\alpha_i \,\delta B_{ij}}{\sum_{k}\alpha_k
        }}^3\right\rangle
        &=
        \pq{\ \left\langle\;\frac{
        \alpha_1^3  }
        {\pt{\sum_{k}\alpha_k
        }^3}\right\rangle
        -
        3\,\left\langle\;\displaystyle\frac{
        \alpha_1^2 \,\alpha_2 }
        {\pt{\sum_{k}\alpha_k
        }^3}\right\rangle
        +
        2\,\left\langle\;\displaystyle\frac{
        \alpha_1 \,\alpha_2 \,\alpha_3}
        {\pt{\sum_{k}\alpha_k
        }^3}\right\rangle\ }
        \ \displaystyle\sum_{\mu}\pt{\delta B_{\mu j}}^3
        \,.\nonumber
        \intertext{Now we use~\eqref{eq75}}
        \left\langle\;\pt{\frac{\sum_{i}\alpha_i \,\delta B_{ij}}{\sum_{k}\alpha_k
        }}^3\right\rangle
        &=
        \pg{\ \frac{\Gamma\pt{\sigma+3}}{{}^{\phantom{2}}\Gamma\pt{\sigma}^{\phantom{2}}}
        -
        3\,\frac{\Gamma\pt{\sigma+2}\Gamma\pt{\sigma+1}}{\pq{\Gamma\pt{\sigma}}^2}
        +
        2\,\pq{\frac{\Gamma\pt{\sigma+1}}{{}^{\phantom{2}}\Gamma\pt{\sigma}^{\phantom{2}}}}^3 }
        \;\frac{\Gamma\pt{N\sigma}}{\Gamma\pt{N\sigma+3}}\;\displaystyle\sum_{\mu}\pt{\delta B_{\mu j}}^3
        \nonumber\\
        &=\frac{2}{N\pt{N\sigma+1}\pt{N\sigma+2}}\;\displaystyle\sum_{\mu}\pt{\delta B_{\mu j}}^3
        \,.\label{eq98}
        \end{align}
Thus, from~\eqref{eq95}\,, the third order contribution to
$\widetilde{P}_{s}\big[\{B_{ij}\}\big]$ is
        \begin{equation}
        \Delta_3 \widetilde{P}_{s}\big[\{B_{ij}\}\big]
        =P^{\star}_N\pg{\frac{\pt{s-1}N^3}{3}-
        \frac{2\,N^3 s}{3\pt{N\sigma+1}\pt{N\sigma+2}}\;}\sum_{pq}
        {\pt{\delta B_{pq}}}^3\label{eq99_1}
        \end{equation}
and, near the center $B^{\star}_N$\,,
$\widetilde{P}_{s}\big[\{B_{ij}\}\big]$ has the following structure:
        \begin{equation}
        \widetilde{P}_{s}\big[\{B_{ij}\}\big]
        =P^{\star}_N\exp\pg{-c_2\sum_{pq}{\pt{\delta
        B_{pq}}}^2-c_3\sum_{pq}
        {\pt{\delta B_{pq}}}^3+
 {\cal O}\pq{\pt{\delta B}^4}}\,.\label{eq99_2}
        \end{equation}
Assuming that $s_i=s$ (so $\sigma=Ns$) we may then find from~(\ref{eq89}) the value of the constant $c_2$,
        \begin{subequations}\label{eq100}
        \begin{align}
 c_2&=1-\frac{N^2}{2}+\frac{\sigma^2\,N^2}{2\pt{\sigma\,N+1}}\,.\label{eq100_2}
 \intertext{Similarly eq.~(\ref{eq99_1}) implies that the third constant reads,}
 c_3&=\frac{N^2}{3}\pt{N-\sigma+
        \frac{2\,\sigma}{\pt{N\sigma+1}\pt{N\sigma+2}}}\,.\label{eq100_3}
        \end{align}
        \end{subequations}
        Adjusting $s=\sigma/N$ appropriately to the size $N$ of the matrix one may find such a value of the Dirichlet parameter
        $s$ that $c_2$ or $c_3$ are equal to zero. However, if we set $c_2$ to zero, the parameter $c_3$ is non zero, so the
        third order terms remain in eq.~(\ref{eq99_2}). Thus we have shown that it is not possible to find
        an initial Dirichlet distribution which gives the output distribution uniform in the vicinity of the center of the
        Birkhoff polytope up to the third order.
        A power expansion of $c_3$ gives
        \begin{equation}
        c_3
        =\frac{N}{3}+\frac{1}{N}-\frac{4}{3}\pt{\frac{1}{N}}^3+
 {\cal O}\pq{\;\pt{\frac{1}{N}}^5}\,,\label{eq101}
        \end{equation}
Thus the scale of the asymmetry is $\delta B\propto N^{-1/3}$
so it cannot be seen for $\abs{\delta B}\lesssim N^{-1/3}$\, that
means if $\abs{\delta B/B^{\star}_N}\lesssim N^{2/3}$\,.

\end{document}